# Evidence for increasing frequency of extreme coastal sea levels

## Authors
Tony E. Wong[1], Travis Torline[2], Mingxuan Zhang[3]

## Affiliations
1. School of Mathematical Sciences, Rochester Institute of Technology, Rochester, NY 14623
2. Department of Computer Science, University of Colorado Boulder, Boulder, CO 80309
3. Department of Statistics, Purdue University, West Lafayette, IN 47907
corresponding author: Tony Wong (aewsma@rit.edu)

## Abstract
Projections of extreme sea levels (ESLs) are critical for managing coastal risks, but are made complicated by deep uncertainties. One key uncertainty is the choice of model structure used to estimate coastal hazards. Differences in model structural choices contribute to uncertainty in estimated coastal hazard, so it is important to characterize how model structural choice affects estimates of ESL. Here, we present a collection of 36 ESL data sets, from tide gauge stations along the United States East and Gulf Coasts. The data are processed using both annual block maxima and peaks-over-thresholds approaches for modeling distributions of extremes. We use these data sets to fit a suite of potentially nonstationary extreme value models by covarying the ESL statistics with multiple climate variables. We demonstrate how this data set enables inquiry into deep uncertainty surrounding coastal hazards. For all of the models and sites considered here, we find that accounting for changes in the frequency of coastal extreme sea levels provides a better fit than using a stationary extreme value model.

## Introduction
Projections of the future coastal hazard posed by extreme coastal sea levels are a key component of designing strategies to manage coastal flood risk, but probabilistic estimates of these hazards suffer from a number of intrinsic uncertainties. These uncertainties in coastal hazard estimates lead to uncertainty in the "correct" strategy to manage coastal risk[1,2]. In turn, the uncertainty in flood defense strategy leads to potentially poorer performance of the risk management strategy in terms of economic losses[3], size of inundated area[4] or loss of life[5]. Thus, it is important to quantify how key uncertainties affect the estimated coastal flood height return levels.

The total extreme sea level height has three main components: mean sea level, astronomical tide levels and storm surge. The present work will focus on the total summed tide and storm surge level, termed "storm tide", because this is the total hazard facing coastal zones. This approach also avoids the issue of biases that may arise when attempting to disentangle the tidal and non-tidal components of extreme sea level[6]. For brevity's sake, we will interchangeably refer to the storm tide component as the "extreme sea level" (ESL), though what we are really referring to is the sea level in substantial excess of the mean. This approach runs the risk of missing the compounding effect on flood hazard of multiple drivers of risk[7,8] and will not capture the potentially sizable costs of frequent minor, or "nuisance", flood events[9,10]. With these caveats in mind, this approach of modeling storm tides as a univariate hazard is effective for analyzing this particular key uncertainty, and is important for policy-making, as it represents the actual water level impacting the coastline during a flood event.

There exist multiple approaches for estimating storm tide return levels. These include the joint probability method[11–13], process-based modeling[14,15] and statistical modeling[16]. A key strength of process-based modeling is that explicitly resolving individual storms and physical processes permits an evaluation of the physical drivers of risk and their spatiotemporal dependence. However, process-based modeling comes at a higher computational cost. Joint probability method, by contrast, models the probabilities associated with flood events by estimating the conditional probabilities for the dependency relationships within the statistical model structure. While these other methods for estimating the hazards posed by coastal ESLs have their respective strengths, at present we restrict our attention to statistical modeling. Statistical modeling is another common and less computationally demanding approach for estimating the hazards posed by ESL. We elect to use statistical modeling because the focus of this work is on an accounting of uncertainty, and statistical modeling approaches enable a more detailed quantification of uncertainty than is typically possible in more computationally expensive direct physical modeling approaches.

Statistical modeling involves estimating the probability distribution of ESL heights from some observational data set that has been processed in such a way that it fulfills the assumptions of the fitted distribution (e.g., independent samples). Common choices for distributions include generalized extreme value distributions (GEV) and generalized Pareto distributions (GPD). A GEV is the limiting distribution for a sequence of independent block maxima, so it is a natural choice for sequences of sea-level extremes. However, GEV models are limited in that they do not resolve temporal variability or changes within a given time block. GPD models, on the other hand, give the distribution of exceedances of a given threshold height. These peaks-over-thresholds models are typically coupled to a Poisson process that gives the rate of threshold exceedances per block of time. While of course no model can be expected to be a perfect representation of the real-world distribution of storm surge hazard, the literature is split on which distribution, GEV or GPD, provides the best mathematical representation of storm surge[17]. Thus, it is prudent to examine the impacts of model structural choice when making estimates of future coastal hazards.

Potential nonstationarity in the hazards posed by coastal storm surges presents another key uncertainty that is receiving an increasingly large share of attention in current research[18,19]. Depending on geographic location, uncertainties in future ESLs may be within the range of natural variability and it can be difficult to say conclusively that storm surge statistics are changing[20]. However, failure to account for potential nonstationarity (if it is in fact present) can result in underestimation of storm surge return levels[21], leading to under-protection of coastal areas. In efforts to address this research need, recent work has examined the timescale on which nonstationary storm surge behavior can be confidently detected[22], model averaging approaches to quantify the degree of belief that can be placed on nonstationary model structures[21], and quantification of indicators of changing ESL behavior[23].

Potential nonstationarity can be incorporated into statistical modeling for storm surge return levels by allowing the parameters of the given distribution to covary with some climate conditions and/or indicator(s) (e.g., [24–26]). For example, previous studies have covaried ESL behavior with global mean surface temperature[22,27,28], global mean sea level[29,30], the North Atlantic Oscillation (NAO) index[25,26] and time (i.e., a linear change)[28]. The work of ref. [21] compared estimates for return levels using as a nonstationary covariate the winter mean (DJF) NAO index, global mean surface temperature, global mean sea level and a simple linear change with time. Here, we examine which of the covariates considered by ref. [21] yields the best-fitting extreme value distributions for a set of tide gauge stations along the United States East and Gulf Coasts, and how the amount of data available affects the choice of best covariate. We divide our set of 36 tide gauge sites into "long" and "short" records by separating the set into groups with more or less than 55 years of data available (which is the median record length). There are no stations with exactly 55 years of data, so each group contains 18 stations. Our use of the terms "long" and "short" are relative, and only meant to be used within the context of this study.

Furthermore, the notion of which covariate or nonstationary model structure fits the data the "best" depends on what goodness-of-fit metric one uses to compare the candidate model structures. Here, we present main results using the negative log-posterior score (NPS; see Methods), but we also include results using the negative log-likelihood (NLL), the Akaike information criterion[31] (AIC) and the Bayesian information criterion[32] (BIC) (see Methods and Supplementary Information).

The inherent dearth of data on environmental and climate extremes can lead to poorly constrained estimates of storm tide return levels. Constraint on uncertain model parameters worsens as the number of parameters that must be estimated increases. Consequently, the degree of confidence in those more complex model structures is reduced[21] (c.f., Figure 1 from ref. [21]). This paucity of observational data to constrain extremes frequently leads to a reliance on simpler nonstationary models with fewer parameters. For example, following this principle of parsimony, Rashid et al.[23] consider only potential nonstationarity in the location parameter for the GEV distribution, which governs the center of the distribution. Among the model structures considered by Lee et al.[27], any nonstationarity in the scale parameter, which governs the width of the GEV distribution, is conditioned on the location parameter already being nonstationary. The model structures of Wong et al.[26] suffer from a similar limitation, but in the context of the GPD class of models. However, changes to only the scale parameter of either a GEV distribution or GPD can also bring about changes to the median for these distributions. Additionally, increases to the scale parameter increase the width of the distribution, leading to an increase in the high-risk upper tails. In short, nonstationarity in either the location or scale parameter (or both) could bring about an increase in flood hazard, but previous research has neglected model structures in which only the scale parameter is nonstationary. Thus, it is important to fully examine how the selection of potentially nonstationary statistical model structure affects the estimated storm tide return levels and associated flood hazard. Toward this end, we fit eight candidate model structures for each of the GEV and GPD families of models (see Methods). Among these structures, we vary the model parameters that are considered to be potentially nonstationary.

We analyze two questions: First, what is the best-fitting form of potentially nonstationary extreme value statistical model for each site? We hypothesize that sites with shorter available data records will find better fits in extreme value models with fewer parameters. Second, we ask: how does uncertainty in the model structural choice affect the range of estimated storm tide return level? We expect that for some sites, the projected change in return level in nonstationary models will be similar in magnitude to the range in estimated changes across the set of candidate models[20]. Furthermore, in those cases, we hypothesize that best-fitting extreme value model for storm tide will be conservative, and the projected change in return level will be at the low end of the set of projections from the suite of models.

We address these questions by examining all possible combinations of nonstationary/stationary statistical model parameters for both GEV and GPD model structures, and for each of the four candidate covariates used by ref. [21]. We evaluate the impacts of the model structural choice on goodness-of-fit for data and estimates of current and future storm tide return levels. Finally, in order to facilitate further inquiry into extreme value statistical modeling for storm tides along the U.S. East and Gulf Coast regions, we provide processed versions of the tide gauge data for each site, the candidate covariate time series, and the optimized parameter values for each candidate distribution for each site. Differences in the processing of tide gauge data can affect the resulting estimates of storm tide hazard[26,29], so making these data available is an important step to maintain consistency among studies. It is our aim that by providing the processed data sets, analysis codes and resulting goodness-of-fit metrics - across many sites and potential model structures - these data sets will provide a pathway for others to examine this multifaceted problem of deep uncertainty surrounding the statistical modeling of coastal flood hazards. The results presented here are not intended to be the final word about the "best" statistical model to capture ESL statistical behavior, but rather to facilitate debate and encourage further inquiry.

## Results

### Covariate choices

When the number of model parameters is not penalized (NLL), all sites prefer some form of nonstationary model structure for both the GEV (Fig. 1a-d) and GPD families of models (Fig. 1e-h). For the GEV distribution, when either NLL or AIC is used for model selection, global mean surface temperature emerges as the most popular covariate choice for sites with short data records, while long-data sites are generally split between the sea level and NAO index covariates (Fig. 1a-b). For the GPD family of models, results are mixed; a sea level covariate is preferred by long-data sites much more frequently than by short-data sites across all goodness-of-fit metrics. Not surprisingly, if the number of model parameters is penalized by using the BIC for model selection, a stationary extreme value model fits best for both the GEV and GPD families (Fig. 1c and g).

The essence of a Bayesian modeling approach is to update our *a priori* beliefs about a model and its parameters in light of the available tide gauge data. The negative (log-)posterior score combines both these prior probabilities and the log-likelihood of the data, given the supposed values for model parameters. Taking the Bayesian approach and using NPS leads to the temperature covariate as the overall favorite (Fig. 1d and h). For the remainder of the analysis presented in the main text, we provide results using NPS for model selection, but analogous results using the other goodness-of-fit measures may be found in the Supplementary Material accompanying this work.

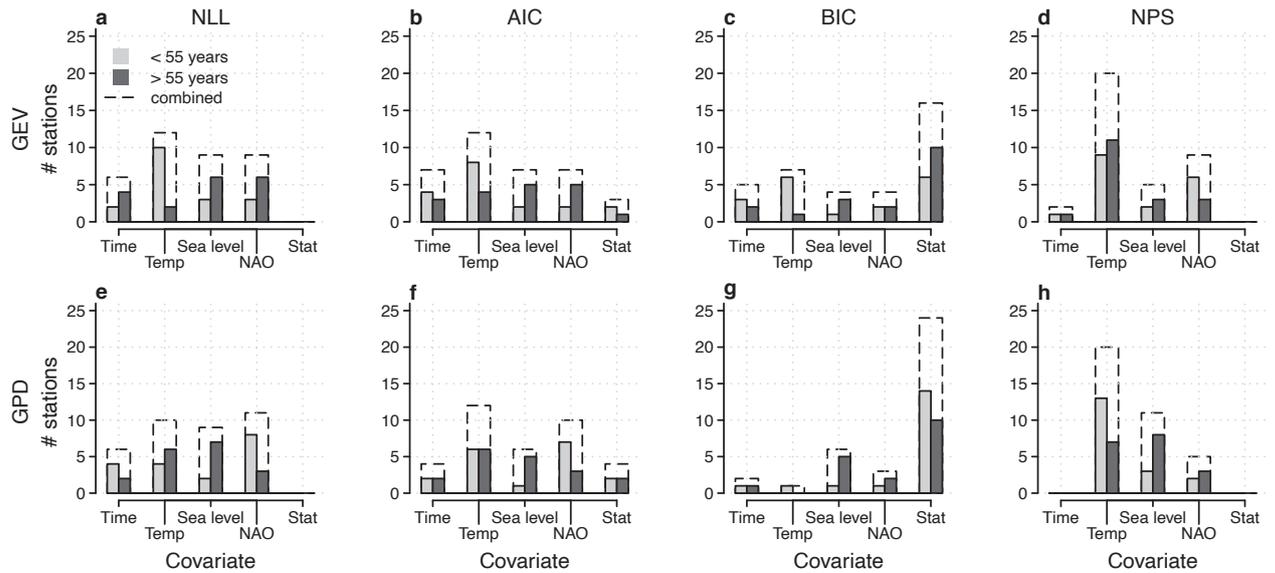

*Figure 1.* Frequency graphs of covariate choice. These figures aggregate over all 36 tide gauge stations, for each of the goodness-of-fit metrics: negative log-likelihood (first column), Akaike information criterion (second column), Bayesian information criterion (third column) and negative posterior score (fourth column). The top row corresponds to consideration of the eight GEV model structures and the bottom row corresponds to the eight GPD models. Sites are separated into long (>55 years, dark gray) and short (<55 years, light gray) data record lengths.

The lack of a clear overall pattern to the choice of covariate by data length raises the question of whether a geographical pattern to these model structural choices might exist. The Gulf Coast shows some preference for all of the covariates at different locations when we use a GEV distribution (Fig. 2a). When we use a GPD, however, the temperature and NAO index covariates are preferred on the western Gulf Coast, and temperature and sea level are the preferred covariates on the eastern Gulf Coast (Fig. 2b). In the mid-Atlantic, the region roughly bounded to the south by the Chesapeake Bay and to the north by New York City shows preference for either a stationary extreme value model or one where nonstationarity is modulated by global temperature. In the Northeast (New England, east and north of New York City), the NAO index and sea level covariates are preferred when using the GPD family of models, but there is no clear trend when using the GEV family.

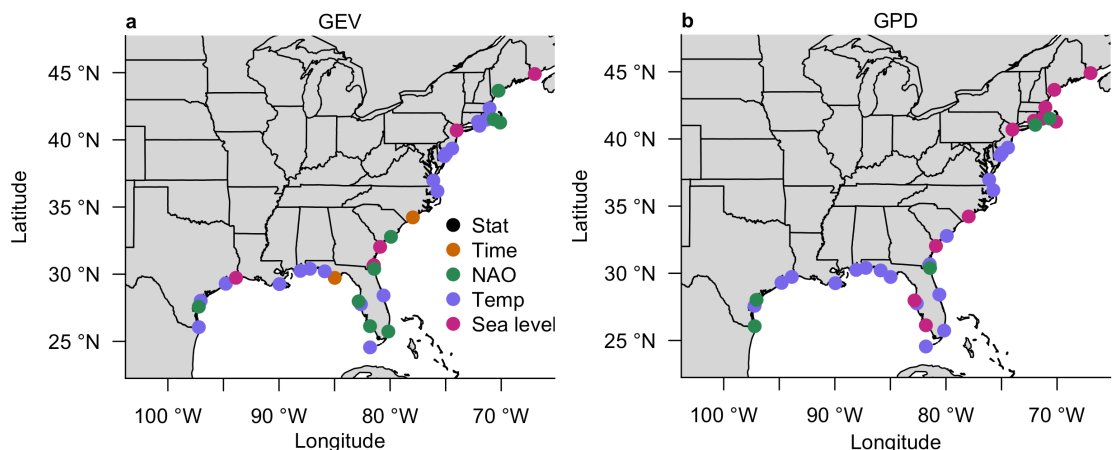

*Figure 2.* Geographical distribution of the preferred covariate time series. Best-fit models are selected using negative log-posterior score to evaluate goodness-of-fit, assuming a GEV model (a) or a GPD one (b).

## Model choice

When we select the best-fitting model using NPS, no sites prefer a GEV model with only the location parameter ($\mu$) nonstationary; if a single GEV model parameter is to be nonstationary, it should be the scale parameter ($\sigma$) (Fig. 3, left column). Several sites with long data records prefer the GEV model wherein both $\mu$ and $\sigma$ are nonstationary, which highlights the strict requirements placed on available data in order to fit more complex nonstationary extreme value models. For a GPD model, the preferred model structures are models with nonstationarity in both the Poisson rate parameter ($\lambda$) and $\sigma$, and nonstationarity in all parameters (Fig. 3, right column). If a site prefers nonstationarity in all three GPD model parameters, then it is more likely to be a site with a shorter data record than a long one (with the exception of the NAO index covariate). Conversely, a site that prefers the GPD model structure with $\lambda$ and $\sigma$ nonstationary is more likely to have a long data record. We take this result to indicate the benefits of more available data to reject potentially unphysical or poorly constrained models with shape parameter ($\xi$) nonstationary (see the discussions in refs. [16] and [22]).

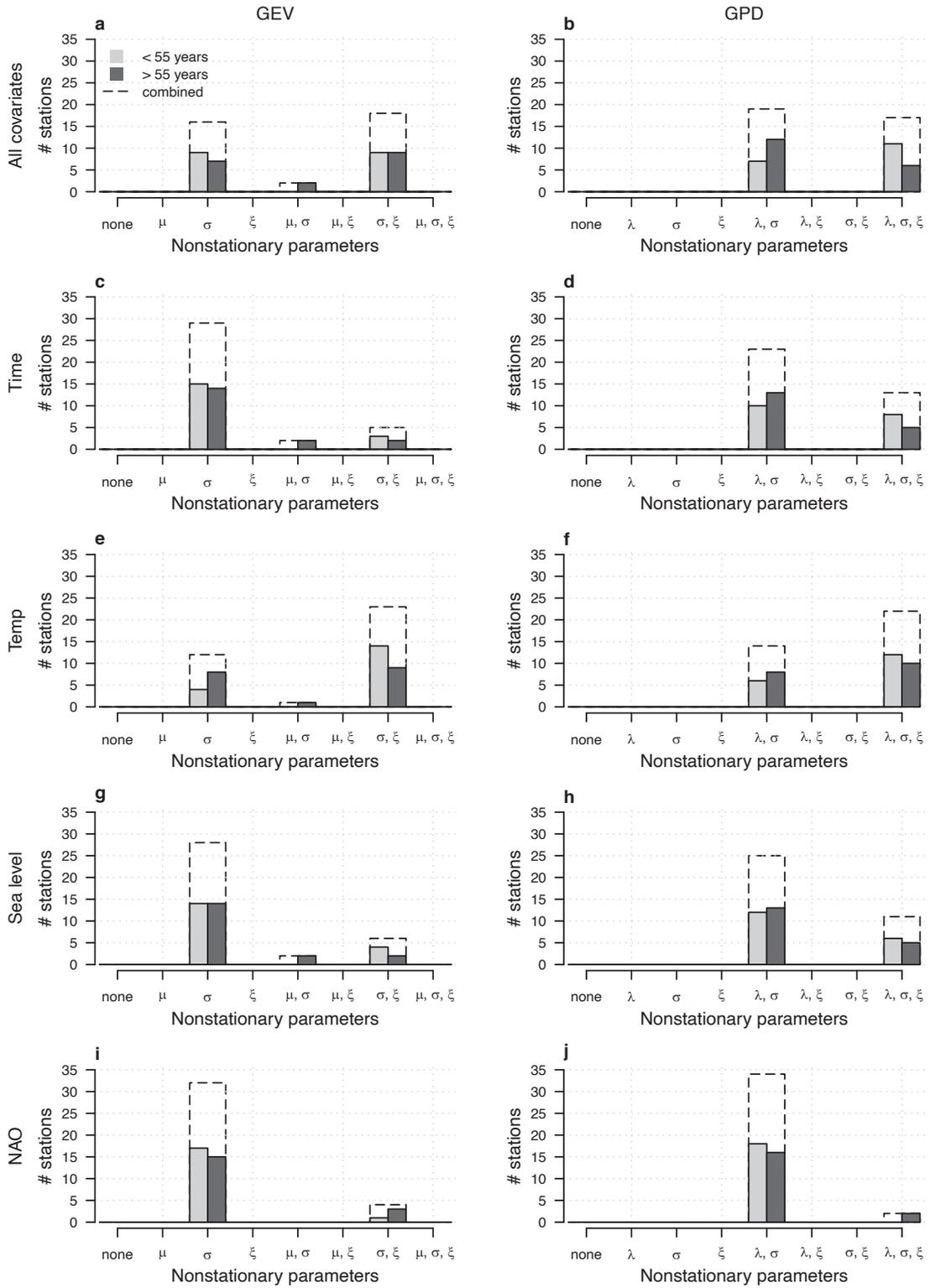

*Figure 3.* Frequency graphs of model choice. These figures aggregate over all 36 tide gauge stations, and over all covariates (top row), or considering each covariate individually (time, second row; temperature, third row; sea level fourth row; and NAO index, fifth row). Left column corresponds to consideration of the eight GEV model structures and right column corresponds to the eight GPD models. Sites are separated into long (>55 years, dark gray) and short (<55 years, light gray) data record lengths. Negative log-posterior score is used to evaluate model goodness-of-fit.

We aggregate all of the covariates, which amounts to equal weighting of all of the candidate covariates and choices for nonstationary parameters (Fig. 3a-b). This offers a potentially cleaner comparison of the different choices of nonstationarity in model parameters. This approach will, of course, miss differences in how the covariate choice affects choice of nonstationary parameters. For example, a poor choice of covariate would likely result in selection of a stationary model structure. However, none of the covariates employed here lead to a stationary model choice when using NPS as the goodness-of-fit measure. Keeping this caveat in mind, we find that GEV models are generally split between nonstationarity in the scale parameter ($\sigma$) only, and nonstationarity in the scale and shape parameters ($\sigma$ and $\xi$). Given the discussions in refs. [16] and [22] about the shape parameter, this result strongly supports the use of a GEV model in which scale ($\sigma$) is the only nonstationary parameter. This model structure strikes a balance between allowing the data to drive the model structural choice and parsimonious use of additional model parameters.

Similarly, the GPD models show a preference for the $\lambda$- and $\sigma$-nonstationary model and the fully nonstationary model (Fig. 3b). Again, considering the arguments against nonstationarity in the shape parameter, we take this as an indication of the strength of the $\lambda$- and $\sigma$-nonstationary GPD model structure. Note, however, that these results do not necessarily suggest that the $\lambda$- and $\sigma$-nonstationary model would be the second-best choice for the models that selected the fully nonstationary GPD model.

## Estimated return levels

In light of the fact that the GPD model can constrain nonstationary model structures with more parameters (c.f. Fig. 3), and the arguments against nonstationarity in $\xi$, we use all four potentially nonstationary GPD model parameter structures with stationary $\xi$ and all four candidate covariates to fit a total of 13 distinct models for each tide gauge data site (4 covariates × 3 nonstationary models + 1 stationary model). We use these models to project the 50-year storm tide return level for each of the sites with at least 55 years of available data (Fig. 4). In the Supplementary Material, we include analogous figures when the GEV family of models is used, and for other return periods.

These results demonstrate the diversity in estimates of flood hazard estimates attributable to model structural uncertainty. As expected, for some sites, the signal is roughly equal in magnitude to the noise. For example, in Key West, Florida and in Boston, Massachusetts, the storm tide by the year 2050 is estimated to be roughly 950 mm and 3100 mm, respectively, but their ranges between the minimum and maximum model projections are 920 mm and 4370 mm, respectively. Not surprisingly, the best-fitting models in each case are on the low ends of the sets of projections (Fig. 4a and g). We are not arguing that the range is the appropriate measure of uncertainty, especially with skewed distributions as tend to arise in dealing with extremes. However, the range in model projections does shed light on the magnitude of the uncertainties present, and our intention is to provide a framework for understanding the uncertainties inherent in this necessary enterprise of modeling flood hazards.

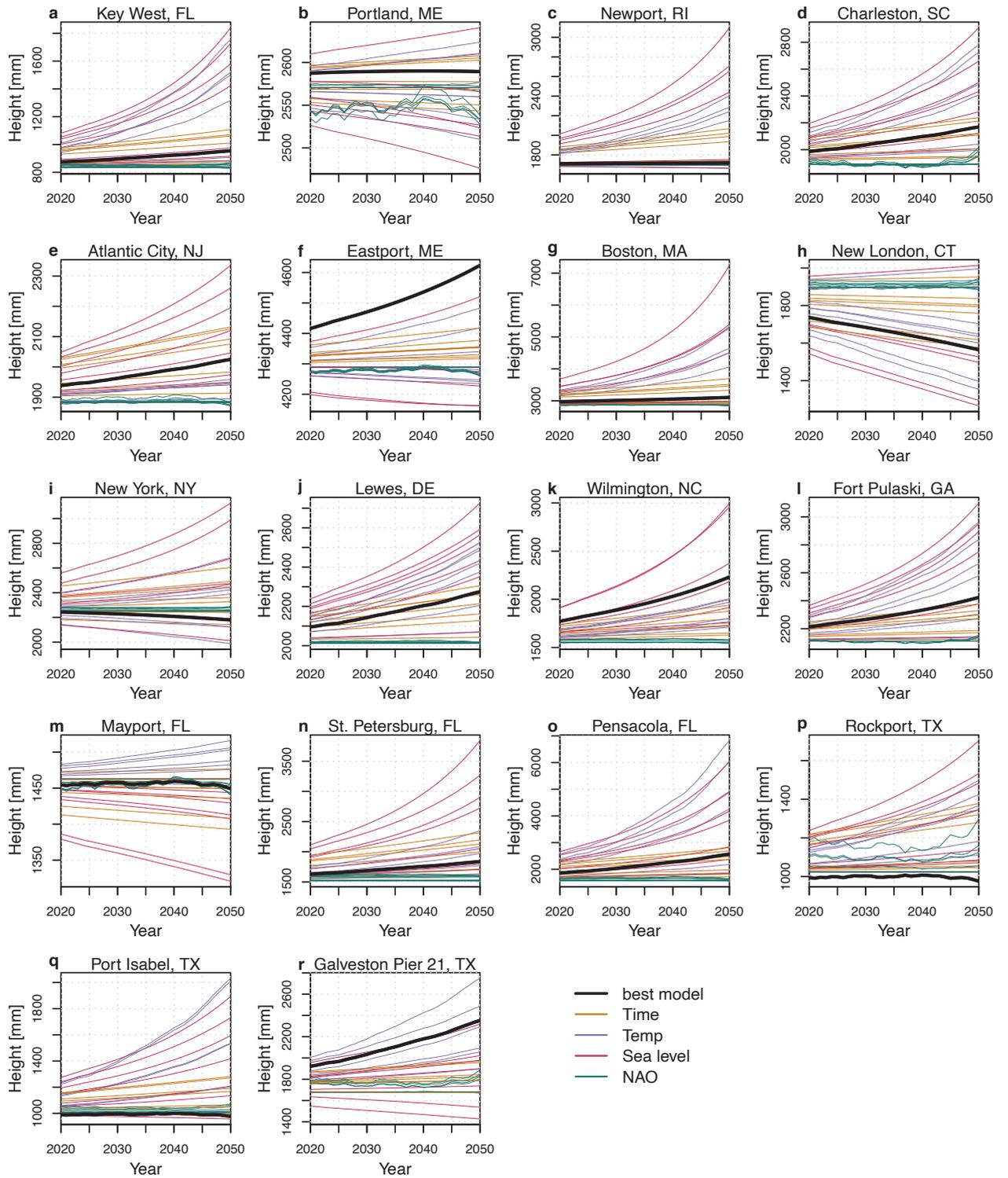

*Figure 4. Projected 50-year return levels for the long-data sites. Uses a GPD model and a centered 21-year moving window average. The best-fitting model is selected by minimizing the negative log-posterior score and is shown in black. The four candidate models using as the covariate time series are shown with the same colored lines: time (orange), temperature (purple), sea level (red) and NAO index (green). There is no visual distinction in each panel among the eight candidate model structures within the set using each covariate.*

## Discussion

We have presented a set of experiments to characterize uncertainty in projected ESL return levels for sites along the U.S. East and Gulf Coasts. In particular, we have focused our attention on three main questions: (i) if we consider a sample of 36 tide gauge stations, which of the candidate covariates and which of the candidate (non)stationary model parametric structures fits best for each site?, (ii) what geographic patterns emerge based on those best-fitting models? and (iii) how does the projected 50-year return level compare to the uncertain range in future return levels, when looking across all of the plausible candidate model structures? We also examined whether the amount of available data has an impact on the choice of best-fitting model structure, with a specific focus on the number of parameters used (see Table 2).

Along the U.S. Gulf Coast and mid-Atlantic regions, global mean surface temperature emerges as the dominant best-fitting covariate for both the GEV and GPD extreme value models (Fig. 2). In the Northeast, winter mean NAO index and global mean sea level also become important covariates for ESL. These geographic differences are likely attributable to the main drivers of ESL in different regions. For example, in the Gulf Coast region, the premise underlying previous work examining ESL in this region assumed a connection between tropical storm activity and changes in surface temperature (namely, in the tropical development region where storms may form)[28]. The fact that temperature is favored as the best-fitting covariate for many of the sites considered here, and for both the GEV and GPD models, supports that premise. We emphasize that these models and methods employ temperature (or sea level, NAO index or time) as a covariate as opposed to a causal mechanism for changes in ESL.

We find that, in contrast to some previous studies[23,26,27], a GEV model with only the scale parameter ($\sigma$) nonstationary fits the annual block maximum data sets better than a model with only the location parameter ($\mu$) nonstationary (c.f., Fig. 3). This result does not undermine these previous works, however. Both model structures lead to a change in the median of the distribution of ESL, but the $\sigma$-nonstationary structure also changes the width of the distribution. An increase in the width as we project return levels further into the future may well accurately reflect our imperfect understanding of precisely how climate change will affect environmental extremes[33,34]. For GPD models, we find that form with nonstationary rate and scale parameters ($\lambda$ and $\sigma$) is the most common best-fitting model structure. When we limit the model choices to only those with a single nonstationary parameter, GPD models with nonstationarity in only the rate parameter are the more popular choice (see Supplementary Material).

Note that we cannot, in general, directly compare goodness-of-fit measures across the GEV and GPD classes of models because the data sets (after processing) employed for the two types of models are different. However, the stationary form of each class of model (GPD or GEV) may be viewed as a baseline against which to compare the nonstationary model structures. For example, the mean differences in NPS among all sites for a global mean sea level, global mean surface temperature or NAO index covariate, using a GPD model with the rate and scale parameters ($\lambda$ and $\sigma$) nonstationary relative to a stationary GPD model, are 11.7, 11.6 and 10.3, respectively. By contrast, the use of a GEV model with only scale ($\sigma$) nonstationary, relative to a stationary GEV model, yields differences in NPS of 5.1, 5 and 4.9, respectively. Given that the use of GPD models constrain two nonstationary parameters and provide a better fit relative to a stationary model, we conclude that a nonstationary GPD model is preferable to a GEV.

Counterintuitively, the sites with shorter records of available data were generally more likely to have a best-fitting GPD model structure with more parameters than the sites with longer data records (Fig. 3, right column). This result, along with the arguments against nonstationary shape parameters[22], suggests that more data is an effective tool to rule out poor choices for model structure. Unfortunately, this is not a tool over which we have much control. However, the GPD model with the Poisson rate parameter $\lambda$ nonstationary is a better fit to all 36 tide gauge sites than a stationary model for all four of the covariates examined here. These results are a strong indictment of the stationary GPD model as inappropriate for projecting storm tide return levels in the U.S. East and Gulf Coast regions.

A critical caveat is that these results for the optimal model choice (by whatever measure one wishes to use) do not depict the strength of the structural choice. For example, the range in AIC for the Chesapeake Bay Bridge Tunnel (Virginia) station for the GEV models is 549.79-550.42. Differences in AIC and BIC this small do not indicate a strong preference among the model structural choices. Thus, there is effectively almost no difference between the best-fitting and the worst-fitting models. However, we include a stationary model in the set of candidate models, so our framework represents the hypothetical decision landscape that faces a modeler tasked with projecting storm tide hazard for these sites: a choice must be made among these (or other) possible model structures. Alternatively, this is an area where model averaging approaches have been shown to be a fruitful avenue to combine projections across model structures[21,26,35], or one might consider selecting a model that minimizes the potential regret if that choice turns out to be inappropriate[36,37].

There are, of course, many other cross-sections of these results that may be of interest to researchers from the perspective of evaluating the practical implications of subjective choices that must be made regarding model structures employed. For example, one might ask: to what degree are estimated return levels sensitive to misspecification of the "correct" model structure for a given site? Or: would model structures that use a local mean surface temperature or local mean sea level covariate time series perform better than the models presented here, which use global mean time series (see discussion in ref. [28])? Indeed, there is even uncertainty and variation between studies in the processing steps used to detrend and ensure independence among extreme sea level events[26,29]. As a step in producing these results, we provide a set of processed tide gauge data for a set of 36 U.S. tide gauge sites. For each site, these data sets can be used to fit GEV distributions to the time series of annual block maxima, or GPD peaks-over-thresholds models. For a GPD model, for example, results are conditioned on the threshold chosen for the peaks-over-thresholds approach and the declustering time-scale for removing multiple extremes that originate from the same event. Both of these processing factors can be adjusted in the code repository that accompanies this work. It is our broader aim that this work may provide a useful framework for these lines of inquiry regarding extreme value statistical models for extreme coastal sea levels, including a common set of processed tide gauge records for two common extreme value modeling approaches.

## Methods
### Data
We use tide gauge data from 36 stations along the United States East and Gulf Coasts, obtained from the University of Hawaii Sea Level Center[38] (accessed 21 March 2020). The site names, locations and dates of available data are given in Table 1. Gaps exist in many of the data sets, which leads to the discarding of some data. Thus, the "Years of usable data" column does not always match the interval between "Start date" and "End date" in Table 1.

| Site | Latitude (°N) | Longitude (°E) | Start date (Y-M-D) | End date (Y-M-D) | Years of usable data |
|---|---|---|---|---|---|
| Dauphin Island, AL | 30.250 | -88.075 | 1981-04-01 | 2018-12-31 | 28 |
| New London, CT | 41.355 | -72.087 | 1938-06-13 | 2018-12-31 | 70 |
| Lewes, DE | 38.782 | -75.120 | 1919-02-02 | 2018-12-31 | 68 |
| Apalachicola, FL | 29.727 | -84.982 | 1976-04-02 | 2018-12-31 | 41 |
| Clearwater Beach, FL | 27.977 | -82.832 | 1996-01-01 | 2018-12-31 | 23 |
| Fernandina Beach, FL | 30.672 | -81.467 | 1897-05-09 | 2018-12-31 | 53 |
| Key West, FL | 24.553 | -81.808 | 1913-01-20 | 2018-12-31 | 102 |
| Mayport, FL | 30.395 | -81.432 | 1928-04-27 | 2000-11-30 | 68 |
| Naples, FL | 26.130 | -81.807 | 1965-03-05 | 2018-12-31 | 47 |
| Panama City Beach, FL | 30.213 | -85.880 | 1993-01-01 | 2018-12-31 | 18 |
| Pensacola, FL | 30.403 | -87.213 | 1923-05-02 | 2018-12-31 | 87 |
| Port Canaveral, FL | 28.415 | -80.593 | 1994-10-22 | 2018-12-31 | 22 |
| St. Petersburg, FL | 27.760 | -82.627 | 1946-12-25 | 2018-12-31 | 69 |
| Virginia Key, FL | 25.732 | -80.162 | 1994-01-29 | 2018-12-31 | 25 |
| Fort Pulaski, GA | 32.033 | -80.902 | 1935-07-02 | 2018-12-31 | 77 |
| Grand Isle, LA | 29.263 | -89.957 | 1980-01-01 | 2018-12-31 | 35 |
| Boston, MA | 42.355 | -71.052 | 1921-05-04 | 2018-12-31 | 95 |
| Nantucket, MA | 41.285 | -70.097 | 1965-02-02 | 2018-12-31 | 48 |
| Woods Hole, MA | 41.523 | -70.672 | 1957-01-02 | 2018-12-31 | 52 |
| Eastport, ME | 44.903 | -66.985 | 1929-09-13 | 2018-12-31 | 72 |
| Portland, ME | 43.657 | -70.247 | 1910-03-05 | 2018-12-31 | 94 |
| Duck Pier, NC | 36.183 | -75.740 | 1978-06-02 | 2018-12-31 | 39 |
| Wilmington, NC | 34.227 | -77.953 | 1935-12-29 | 2018-12-31 | 79 |
| Atlantic City, NJ | 39.355 | -74.418 | 1911-08-20 | 2018-12-31 | 95 |
| Cape May, NJ | 38.968 | -74.960 | 1965-11-22 | 2018-12-31 | 44 |
| Montauk, NY | 41.048 | -71.960 | 1947-09-08 | 2018-12-31 | 53 |
| New York, NY | 40.700 | -74.015 | 1920-06-02 | 2018-12-31 | 72 |
| Newport, RI | 41.505 | -71.327 | 1930-09-11 | 2018-12-31 | 77 |
| Charleston, SC | 32.782 | -79.925 | 1901-01-02 | 2018-12-31 | 94 |
| Corpus Cristi, TX | 27.580 | -97.217 | 1983-12-02 | 2018-12-31 | 31 |
| Galveston Pleasure Pier, TX | 29.287 | -94.790 | 1957-08-22 | 2011-07-19 | 50 |
| Galveston Pier 21, TX | 29.310 | -94.793 | 1904-01-02 | 2018-12-31 | 105 |
| Port Isabel, TX | 26.060 | -97.215 | 1944-04-02 | 2018-12-31 | 63 |
| Rockport, TX | 28.022 | -97.047 | 1937-03-02 | 2018-12-31 | 57 |
| Sabine Pass, TX | 29.730 | -93.870 | 1985-01-29 | 2018-12-31 | 31 |
| Chesapeake Bay Bridge, VA | 36.967 | -76.113 | 1975-01-30 | 2016-12-31 | 41 |

***Table 1.*** *Tide gauge station name, latitude, longitude and years of available data.*

We process the raw tide gauge data by first subtracting the annual block means using a moving 1-year window to account for changes in mean sea level. What remains is attributable to storm tides (astronomical tides and storm surges) and natural variability. For modeling using GEV distributions, we aggregate the detrended data into annual blocks and compute the annual mean for each block. We discard any year that is missing more than 90% of its hourly data. Table 2 gives the final number of annual block maxima remaining for analysis for each tide gauge station.

The GPD family of models requires a time series of exceedances of some appropriately chosen threshold extreme sea level. We follow the same procedure as has been used extensively in previous work, so the interested reader is directed to refs. [17,26,29] for further details and a discussion of the associated uncertainties. For these models, we first detrend the data to account for mean sea-level change as described above. We then compute the time series of daily maximum sea levels for each location. We discard any days that are missing more than 90% of the hourly data (that is, 4 hours or more are missing). For each site, we compute the 99th percentile of these daily maximum sea levels to use as the threshold for the GPD peaks-over-thresholds model. We collect a time series of exceedances of this threshold, which we then decluster in order to remove multiple extremes from the same ESL event. We use a declustering timescale of 3 days[26]. Previous experiments suggest that results are not strongly sensitive to this choice of declustering timescale[26]. The final declustered time series of threshold exceedances serves as the data for analysis and fitting a Poisson process/GPD model.

To account for gaps in the data for the GPD analysis, we use the following procedure. We fit an overall linear trend to the raw tide gauge data and subtract the fitted line. We fit a mean annual cycle to the detrended hourly sea levels. We then re-impose the linear trend and fill any gaps with this mean annual cycle plus a linear trend. We do not use any data that was gap-filled for analysis, only for detrending. Averaging is a smoothing operation, so this procedure will dampen extremes near these gaps. To partially account for this, the Poisson process rate parameter accounts for only the portion of the year of data that is present.

To incorporate nonstationarity by modulating the ESL statistical model parameters, we use the four time series covariates from a recent study[21]: the winter mean (DJF) North Atlantic Oscillation (NAO) index, the global mean surface temperature, the global mean sea level and time. We compute the NAO index from sea-level pressure data following Stephenson et al.[39]. We use the historical monthly NAO index data from Jones et al.[40] for the hindcast calibration and we use the MPI-ECHAM5 sea level pressure projection under the Special Report on Emission Scenarios (SRES) A1B as part of the ENSEMBLES project[41]. We take the historical global mean surface temperatures from the National Centers for Environmental Information data portal[42] and the future projections from the CNRM-CM5 simulation (member 1) under Representative Concentration Pathway 8.5 (RCP8.5) as part of the CMIP5 multi-model ensemble (http://cmip-pcmdi.llnl.gov/cmip5/, last accessed 7 July 2017). We use the historical sea level time series from Church and White[43] and use the future projections of Wong and Keller[2]. We note that more recent sea-level rise projections exist, including those using the same model as ref. [2], but keep this sea level time series for consistency with previous work in this area[21]. Other sea level covariate time series, including local sea-level rise, may be incorporated via modifications to the "get_timeseries_covariates.R" function provided in the code repository accompanying this work. The time covariate is a simple identify function, representing a linear increase over time. Each covariate time series is normalized to the 0-1 range over the hindcast period (the time from the first processed data point to present).

## Statistical models for storm surge

### Generalized extreme value distribution

We consider two potential extreme value distributions: a generalized extreme value (GEV) distribution and a generalized Pareto distribution (GPD). The probability density function (pdf) for the GEV distribution is given by

$$f(w(t) \mid \mu(t), \sigma(t), \xi(t)) = \frac{1}{\sigma(t)} r(t)^{\xi(t)+1} e^{-r(t)}$$

where $w(t)$ is the detrended annual maximum sea level in year $t$; $\mu(t)$ is the location parameter (mm), which governs the center of the distribution; $\sigma(t)$ is the scale parameter (mm), which governs the width of the distribution; and $\xi(t)$ is the shape parameter (unitless), which governs the tail of the distribution. The factor $r(t)$ is given by $r(t) = \left(1 + \xi(t)\frac{w(t)-\mu(t)}{\sigma(t)}\right)^{-1/\xi(t)}$ for $\xi(t) \neq 0$ and $\exp[-(w(t)-\mu(t))/\sigma(t)]$ if $\xi(t) = 0$. We assume that the model parameters can vary with time $t$ and are constant within a given year. The general nonstationary form of the GEV model parameters is

$$\mu(t) = \mu_0 + \mu_1\phi(t)$$
$$\sigma(t) = exp(\sigma_0 + \sigma_1\phi(t))$$
$$\xi(t) = \xi_0 + \xi_1\phi(t),$$

where $\mu_0$, $\mu_1$, $\sigma_0$, $\sigma_1$, $\xi_0$ and $\xi_1$ are all constant parameters that depend on location and $\varphi(t)$ is a covariate time series. For each of the four candidate covariates (time, temperature, sea level and NAO index), we consider eight different possible GEV model structures to account for potential non-stationarity in each of the three parameters, $\mu$, $\sigma$ and $\xi$ (see Table 1). Note that the model structure in which $\mu_0=\sigma_0=\xi_0=0$ corresponds to the assumption of stationarity in the GEV distribution, so there is a total of 29 distinct model structures.

| # | GEV | GPD |
|---|---|---|
| 1 | $\mu, \sigma, \xi$ all stationary | $\lambda, \sigma, \xi$ all stationary |
| 2 | $\mu$ nonstationary | $\lambda$ nonstationary |
| 3 | $\sigma$ nonstationary | $\sigma$ nonstationary |
| 4 | $\xi$ nonstationary | $\xi$ nonstationary |
| 5 | $\mu, \sigma$ nonstationary | $\lambda, \sigma$ nonstationary |
| 6 | $\mu, \xi$ nonstationary | $\lambda, \xi$ nonstationary |
| 7 | $\sigma, \xi$ nonstationary | $\sigma, \xi$ nonstationary |
| 8 | $\mu, \sigma, \xi$ nonstationary | $\lambda, \sigma, \xi$ nonstationary |

*Table 2. Candidate model parametric structures.*

The detrended series of annual block maxima, $w$, is assumed to be independent of one another. So, the likelihood function for the GEV distribution is given by

$$L(w \mid \mu_0, \mu_1, \sigma_0, \sigma_1, \xi_0, \xi_1) = \prod_{i=1}^{N} f(w_i \mid \mu(t), \sigma(t), \xi(t)),$$

where $w_i$ denotes the annual block maximum for year $i$.

**Generalized Pareto distribution**
The generalized Pareto distribution (GPD) model employs a peaks-over-thresholds approach in which the set of exceedances of a threshold, $\mu$, is assumed to follow a generalized Pareto distribution. Following previous work, we take the threshold $\mu$ to be the 99th percentile of the time series of detrended daily block maxima (e.g., [26,44,45]). The interested reader is directed to refs. [29] and [26] for a more detailed discussion of the sensitivities surrounding the choice of this threshold and other structural choices. Here, however, we keep this parameter fixed. The pdf for the GPD is given by

$$f(x(t) \mid \mu, \sigma(t), \xi(t)) = \frac{1}{\sigma(t)}\left(1 + \xi(t)\frac{x(t)-\mu}{\sigma(t)}\right)^{-\left(1+\frac{1}{\xi(t)}\right)},$$

where $x(t)$ is sea level height at time $t$, $\sigma(t)$ is the GPD scale parameter (mm) and $\xi(t)$ is the GPD shape parameter (unitless), both of which are assumed to vary with time $t$, and are constant within a given year. Exceedances of the threshold $\mu$ are assumed to occur following a Poisson process with rate parameter $\lambda(t)$ (units of exceedances per day). Supposing that there are $n(t)$ exceedances in the time interval $[t, t+\Delta t]$, the Poisson probability mass function (pmf) is then

$$g(n(t) \mid \lambda(t)) = \frac{(\lambda(t)\,\Delta t)^{n(t)}}{n(t)!} exp(-\lambda(t)\,\Delta t).$$

Like the GEV model parameters, the general nonstationary form of the GEV model parameters is

$$\lambda(t) = \lambda_0 + \lambda_1 \phi(t)$$
$$\sigma(t) = exp(\sigma_0 + \sigma_1 \phi(t))$$
$$\xi(t) = \xi_0 + \xi_1 \phi(t),$$

where $\lambda_0$, $\lambda_1$, $\sigma_0$, $\sigma_1$, $\xi_0$ and $\xi_1$ are all constant parameters that depend on the site location. We use the same candidate covariate times series $\varphi(t)$ as with the GEV distributions (Sec. 2.2.1), and the eight candidate GPD model structures are analogous to those considered for the GEV distributions, with the exception that the location parameter (GEV) is exchanged for the Poisson rate parameter (GPD) (see Table 2). Combining the pdf for the GPD, after conditioning on the number of threshold exceedances, $n(y_i)$, for each year $i=1, 2, …, N$, the likelihood function for the data set of threshold exceedances $x$, given the model parameter set ($\lambda_0$, $\lambda_1$, $\sigma_0$, $\sigma_1$, $\xi_0$, $\xi_1$) is

$$L(x \mid \lambda_0, \lambda_1, \sigma_0, \sigma_1, \xi_0, \xi_1) = \prod_{i=1}^{N} g(n(y_i) \mid \lambda_0, \lambda_1) \prod_{j=1}^{n(y_i)} f(x_j(y_i) \mid \mu, \sigma(t), \xi(t))$$

where $y_i$ denotes the year indexed by $i$ and $x_j(y_i)$ is the $j$th threshold exceedance in year $y_i$. The second product in this equation is replaced by one for any year with no exceedances. We purposefully use $w$ to denote the set of annual block maxima for the GEV model and a separate variable $x$ to denote the set of threshold exceedances for the GPD model. This distinction makes it clear that the sets of data used to fit each model are different.

## Model calibration

For each site we calibrate each candidate model (both GEV and GPD, for each of the four covariate time series) using a differential evolutionary algorithm[46] to maximize the model likelihood function for the time series of annual block maxima (GEV) or threshold exceedances (GPD), given the model parameters and covariate time series. Each annual block maximum or threshold exceedance, after appropriate declustering and detrending, is assumed to be independent of one another.

## Goodness-of-fit

Let NLL denote the negative of the maximal value of the log-likelihood function for a candidate extreme value model (that is, GEV or GPD, along with the specific parameters considered to be potentially nonstationary). Lower values (more negative) of NLL correspond to better model fits to the available data. Suppose the model has $k$ uncertain parameters that must be estimated as in Sect. 2.3 and the data set for analysis has a total of $N$ data points. Then the Akaike Information Criterion[31] (AIC) and the Bayesian Information Criterion[32] (BIC) are given by the following:

$$AIC = 2k + 2\,NLL$$
$$BIC = k \ln(N) + 2\,NLL.$$

Lower values of AIC and BIC correspond to better model fits to the available data. The three goodness-of-fit metrics, NLL, AIC and BIC, constitute a progressive increase in the penalty exacted for using too many model parameters (in that order). Using the likelihood function alone constitutes no penalty for the number of parameters, and the BIC penalizes each additional model parameter by a factor of ln(*N*). For data sets with more than exp(2) (about 7.4) data points, the AIC penalizes additional model parameters less harshly than the BIC. This is true for all of the data sets presented here. Thus, AIC and BIC embody the principle of parsimonious use of parameters and data.

Another perspective which penalizes profligate parameter usage is to apply a Bayesian approach. The model parameters, $\theta$, are assigned a joint prior distribution, $\pi$, and instead of minimizing the negative log-likelihood, we minimize the negative log-posterior score, which is given by a gentle rearrangement of Bayes' theorem as

$$NPS = -\ln(L(x|\theta)) - \ln(\pi(\theta)).$$

We note that this posterior score is proportional to the actual posterior probability from Bayes' theorem, so minimizing NPS is equivalent to maximizing the posterior probability of the parameters. We fit the prior distributions, $\pi$, by using the following procedure. We obtain and process the raw hourly tide gauge data for 27 sites from the University of Hawaii Sea Level Center[38] (accessed 24 July 2017) data repository. For each candidate model structure, for each site, we estimate maximum likelihood parameters. We pool the maximum likelihood parameters for all 27 sites and fit either a normal distribution or a gamma distribution to the set of parameters, depending on whether the parameter's support is theoretically infinite (e.g., the GEV/GPD shape parameter) or half-infinite (e.g., the GEV/GPD scale parameter). Note that these distributions are only (half-)infinite in theory; in practice, extreme values for these parameters are uncommon. However, the strength of using NPS for model selection lies in the prior distribution's function to penalize extreme parameter values that are unlikely a priori.

To examine the impacts of the amount of available data on the structure of the best-fitting statistical models, we use the median of the number of years of available data (55 years, c.f., Table 1) to separate the tide gauge sites into "long" and "short" records. There are no stations with exactly 55 years of data available, so each group contains 18 stations. Our use of the terms "long" and "short" are relative, and only meant to be used within the context of this study.

## Estimates of storm tide return levels

For each site, we use all of the candidate model structures with stationary shape parameters (Table 2, models 1, 2, 3 and 5), conditioned separately on using a GPD (main text) or a GEV model (see Supplementary Material), to estimate the 50-year return level between 2020 and the year 2050. We use NPS as the goodness-of-fit metric for model selection and highlight the best-fitting model for each of the tide gauge sites with more than 55 years of data available (Fig. 4). To illustrate the model structural uncertainty associated with these storm tide statistical models, we show this best-fit model alongside the other 12 model structures considered in this experiment (4 covariates × 3 nonstationary parametric forms).

## Data Availability

All data supporting this work are freely available from https://github.com/tonyewong/surge_comparison (https://doi.org/10.5281/zenodo.3889897) under the GNU general public open-source license.

## Code Availability
All modeling and analysis codes supporting this work are freely available from https://github.com/tonyewong/surge_comparison (https://doi.org/10.5281/zenodo.3889897) under the GNU general public open-source license. The authors assume no responsibility for the (mis)use of these codes.

## Acknowledgements
We gratefully acknowledge Klaus Keller, Vivek Srikrishnan and Nathan Urban for fruitful conversations. We acknowledge the World Climate Research Programme's Working Group on Coupled Modelling, which is responsible for CMIP, and we thank the climate modeling groups (listed in the accompanying Supplementary Material) for producing their model output and making it available. For CMIP the US Department of Energy's Program for Climate Model Diagnosis and Intercomparison provided coordinating support and led development of software infrastructure in partnership with the Global Organization for Earth System Science Portals. The ENSEMBLES data used in this work were funded by the EU FP6 Integrated Project ENSEMBLES (contract number 505539), whose support is gratefully acknowledged. T.T.'s portion of the work was supported by a UROP grant through the University of Colorado Boulder.

## Author contributions
T.T. conducted the initial analysis, produced figures and contributed to the final version of the paper.
T.W. conceived the study, guided the research, generated the final analysis, produced figures, maintained the model and analysis codes and wrote the initial version of the paper.
M.Z. conducted the initial model calibration experiments, performed final validation experiments and contributed to the final version of the paper.

## Competing interests
The authors declare no competing interests.

# Supplementary Information

Accompanying *Evidence for increasing frequency of extreme coastal sea levels*, by Tony E. Wong, Travis Torline, and Mingxuan Zhang

## 1 Figures and Tables

| Modeling center (or group) | Institute ID | Model name |
|---|---|---|
| Centre National de Recherches Météorologiques/ Centre Européen de Recherche et Formation Avanceé en Calcul Scientifique | CNRM-CERFACS | CNRM-CM5 |

***Supplementary Table 1.*** *CMIP5 models employed in the present study.*

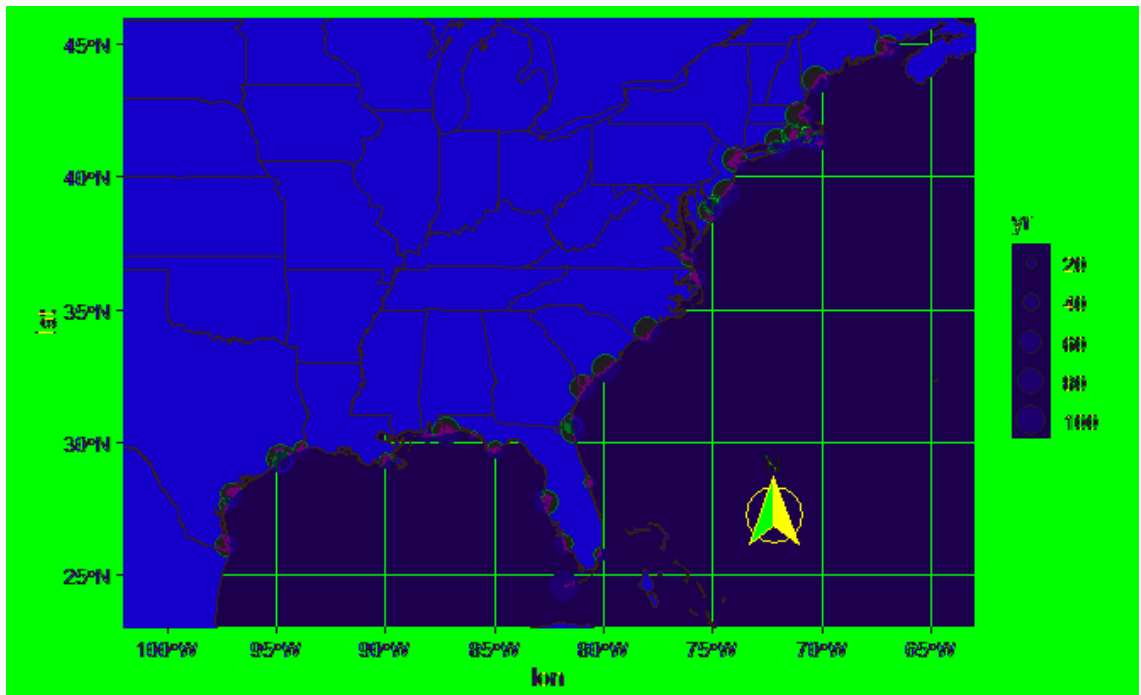

***Supplementary Figure 1.*** *Locations of tide gauge stations used in this work. Dot sizes are proportional to the length of the available data record.*

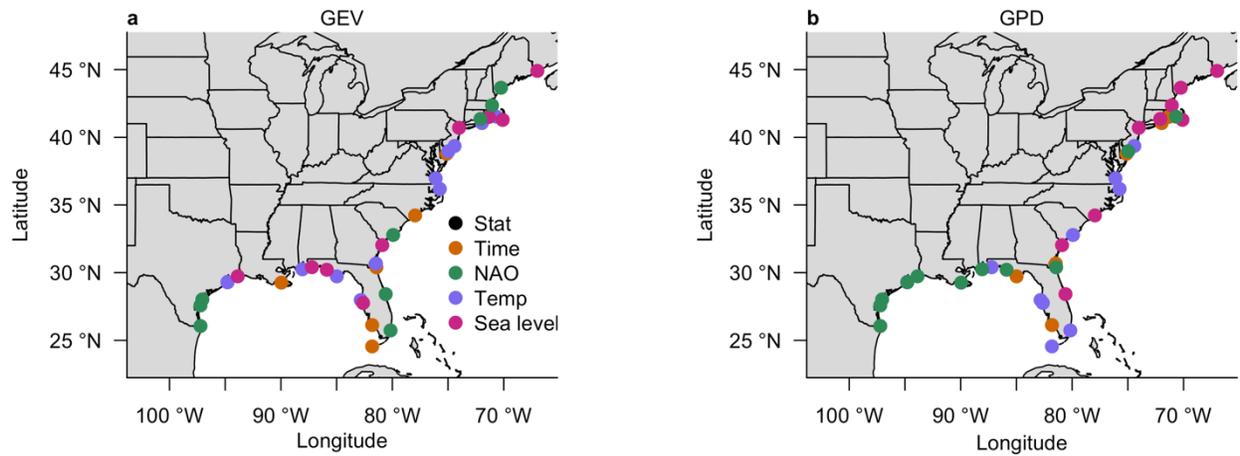

***Supplementary Figure 2.*** *Geographical distribution of the preferred covariate time series. Best-fit models are selected using negative log-likelihood to evaluate goodness-of-fit, assuming a GEV model (a) or a GPD one (b).*

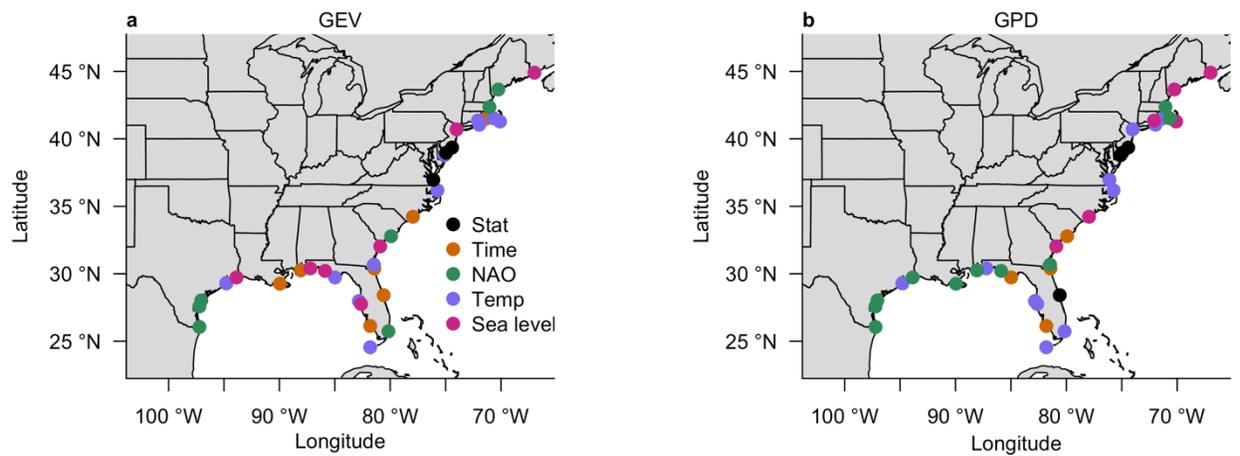

***Supplementary Figure 3.*** *Geographical distribution of the preferred covariate time series. Best-fit models are selected using Akaike information criterion to evaluate goodness-of-fit, assuming a GEV model (a) or a GPD one (b).*

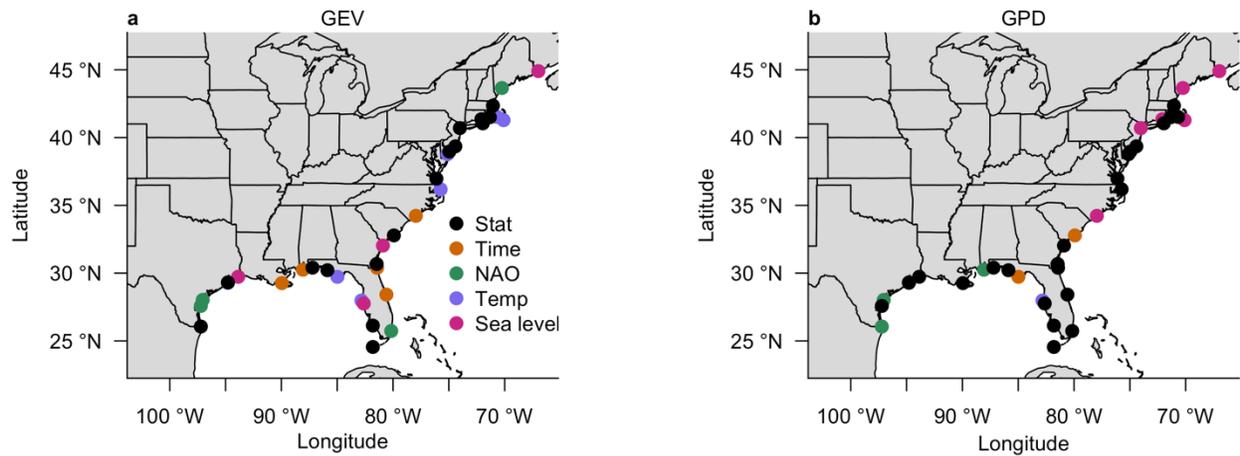

***Supplementary Figure 4.*** *Geographical distribution of the preferred covariate time series. Best-fit models are selected using Bayesian information criterion to evaluate goodness-of-fit, assuming a GEV model (a) or a GPD one (b).*

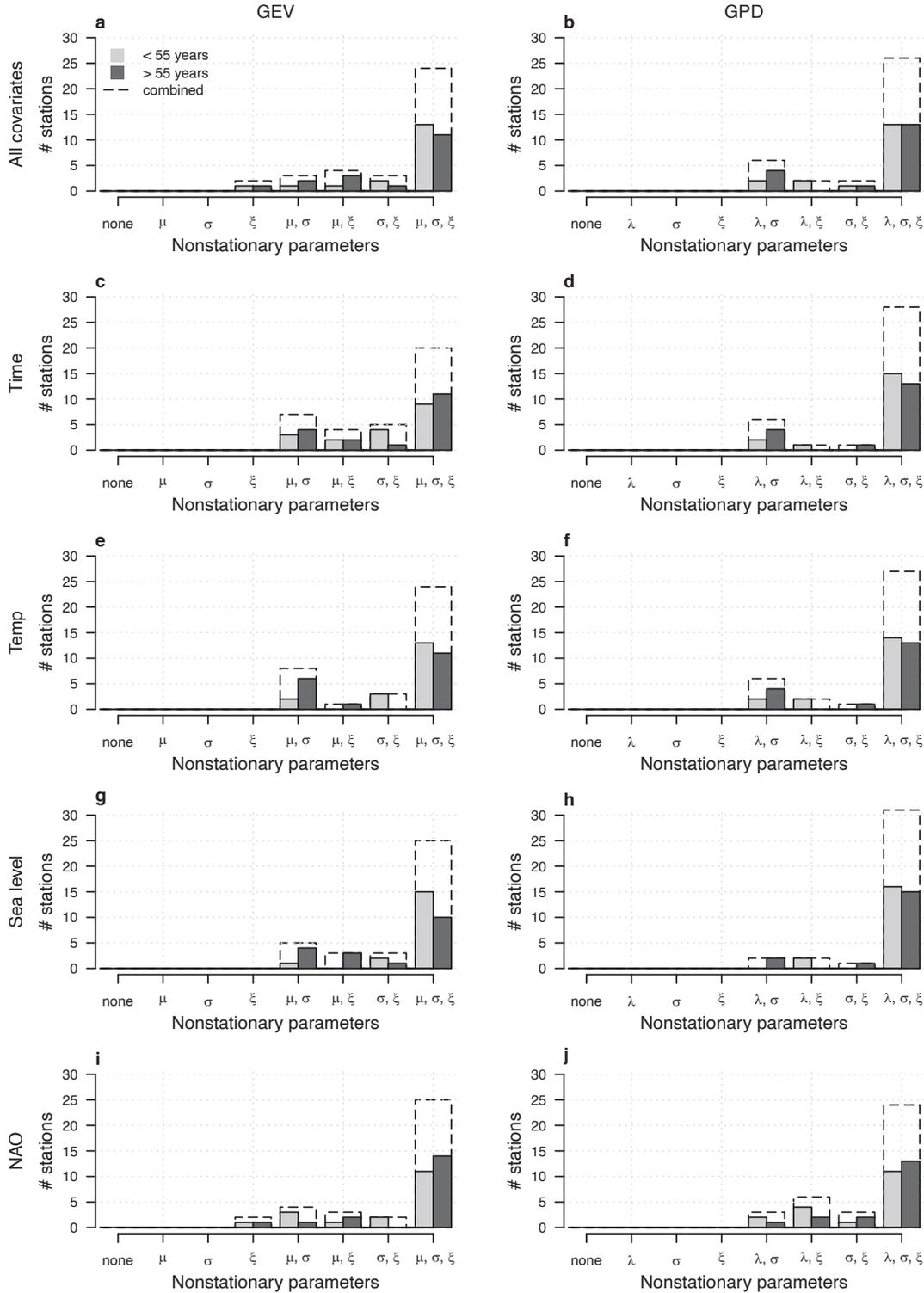

***Supplementary Figure 5.*** *Frequency graphs of model choice. Aggregates over all 36 tide gauge stations, and over all covariates (top row), or considering each covariate individually (time, second row; temperature, third row; sea level fourth row; and NAO index, fifth row). Left column corresponds to consideration of the eight GEV model structures and right column corresponds to the eight GPD models. Sites are separated into long (>55 years, dark gray) and short (<55 years, light gray) data record lengths. Negative log-likelihood is used to evaluate model goodness-of-fit.*

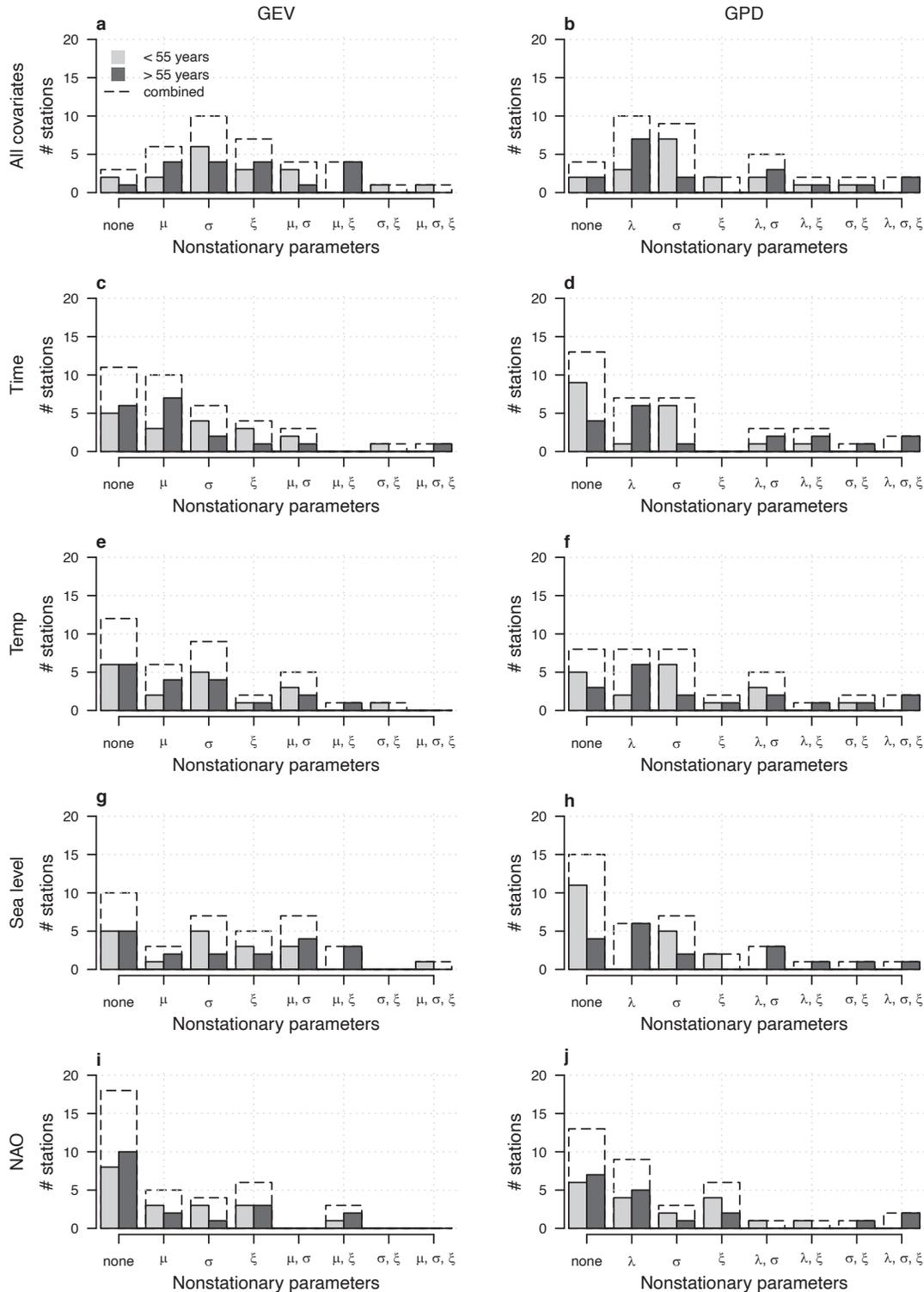

*Supplementary Figure 6.* Frequency graphs of model choice. Aggregates over all 36 tide gauge stations, and over all covariates (top row), or considering each covariate individually (time, second row; temperature, third row; sea level fourth row; and NAO index, fifth row). Left column corresponds to consideration of the eight GEV model structures and right column corresponds to the eight GPD models. Sites are separated into long (>55 years, dark gray) and short (<55 years, light gray) data record lengths. Akaike information criterion is used to evaluate model goodness-of-fit.

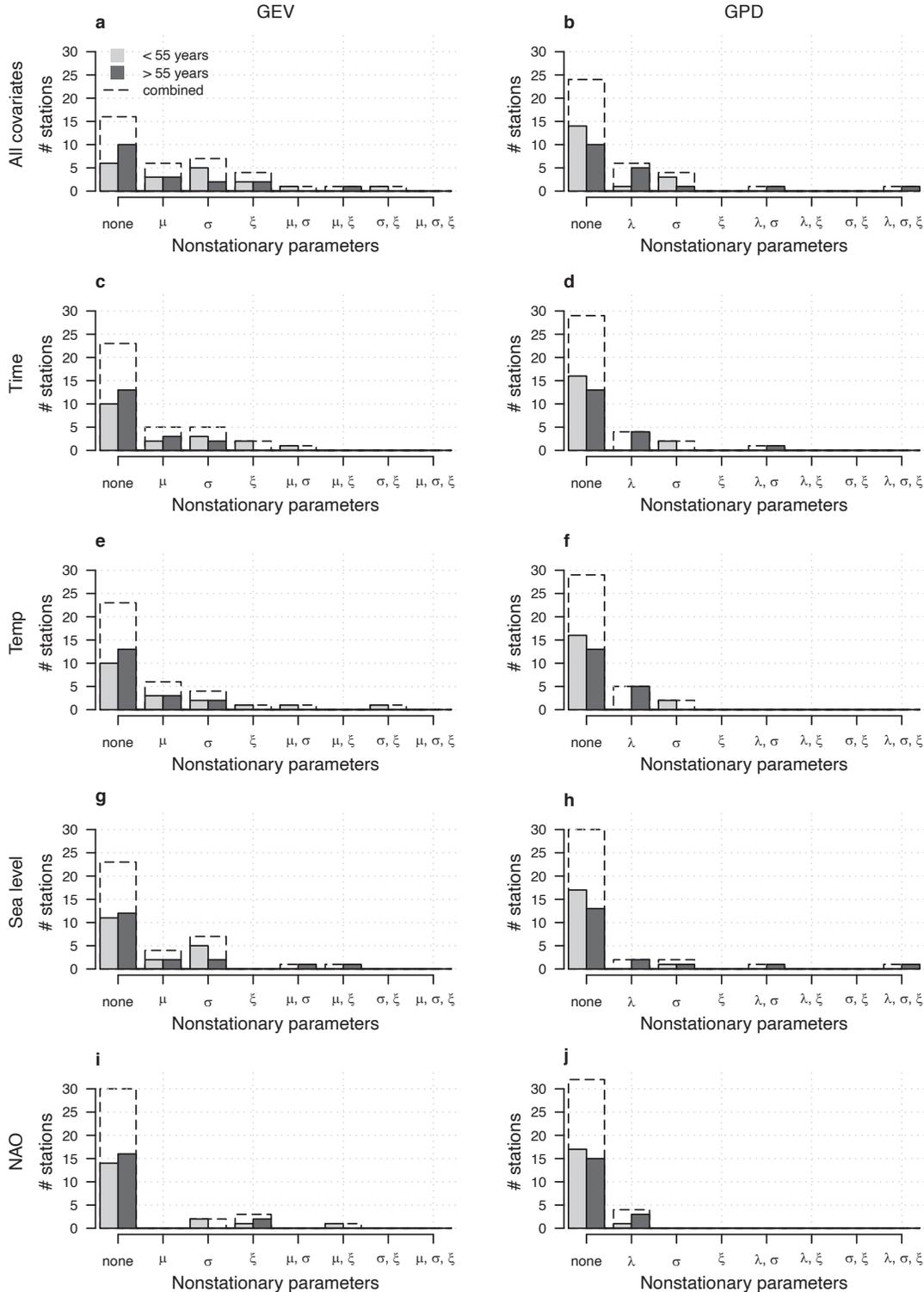

***Supplementary Figure 7.*** *Frequency graphs of model choice. Aggregates over all 36 tide gauge stations, and over all covariates (top row), or considering each covariate individually (time, second row; temperature, third row; sea level fourth row; and NAO index, fifth row). Left column corresponds to consideration of the eight GEV model structures and right column corresponds to the eight GPD models. Sites are separated into long (>55 years, dark gray) and short (<55 years, light gray) data record lengths. Bayesian information criterion is used to evaluate model goodness-of-fit.*

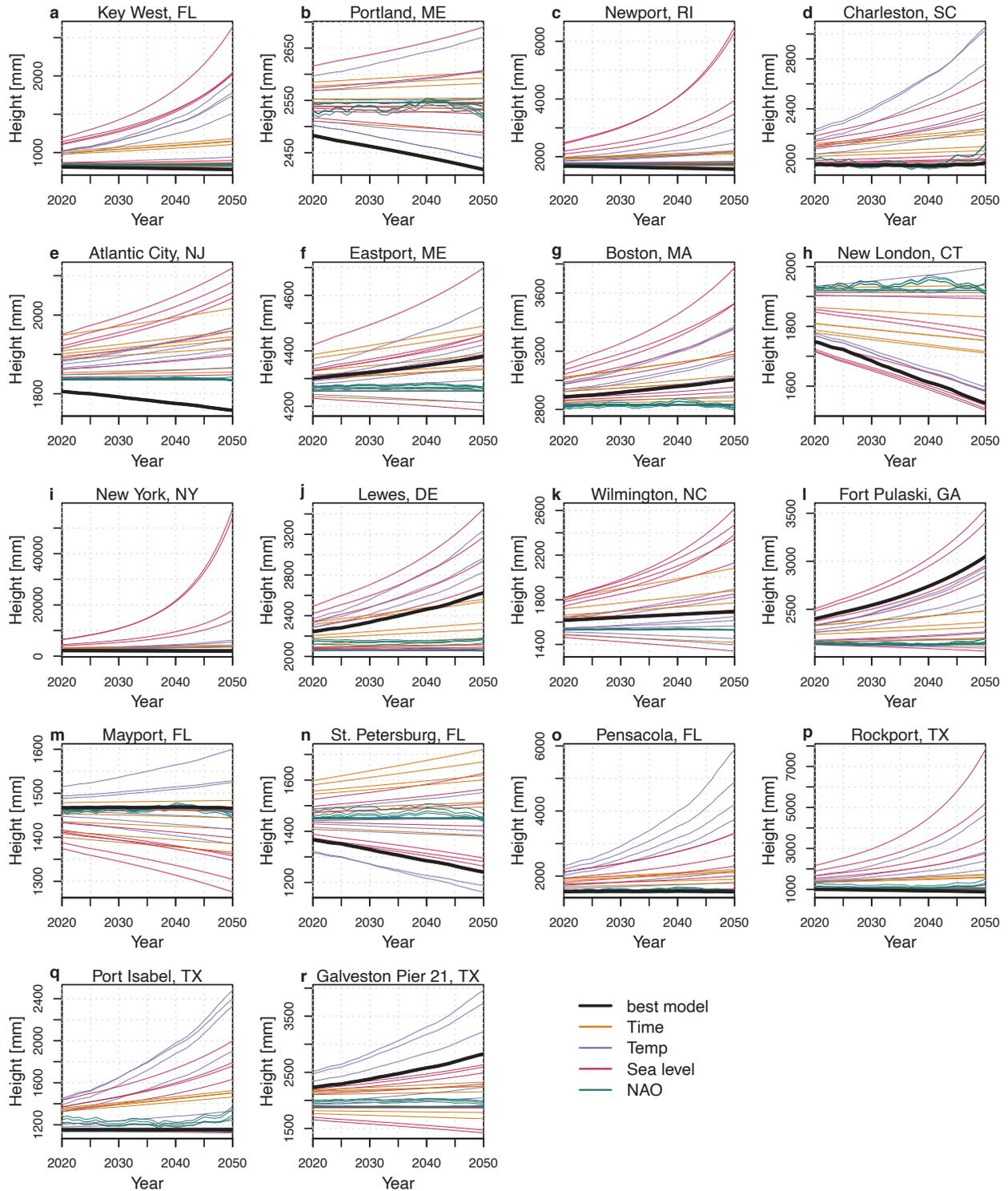

***Supplementary Figure 8.*** *Projected 50-year return levels for the long-data sites. Uses a GEV model and a centered 21-year moving window average. The best-fitting model is selected by minimizing the negative log-posterior score and is shown in black. The four candidate models using as the covariate time series are shown with the same colored lines: time (orange), temperature (purple), sea level (red) and NAO index (green). There is no visual distinction in each panel among the eight candidate model structures within the set using each covariate.*

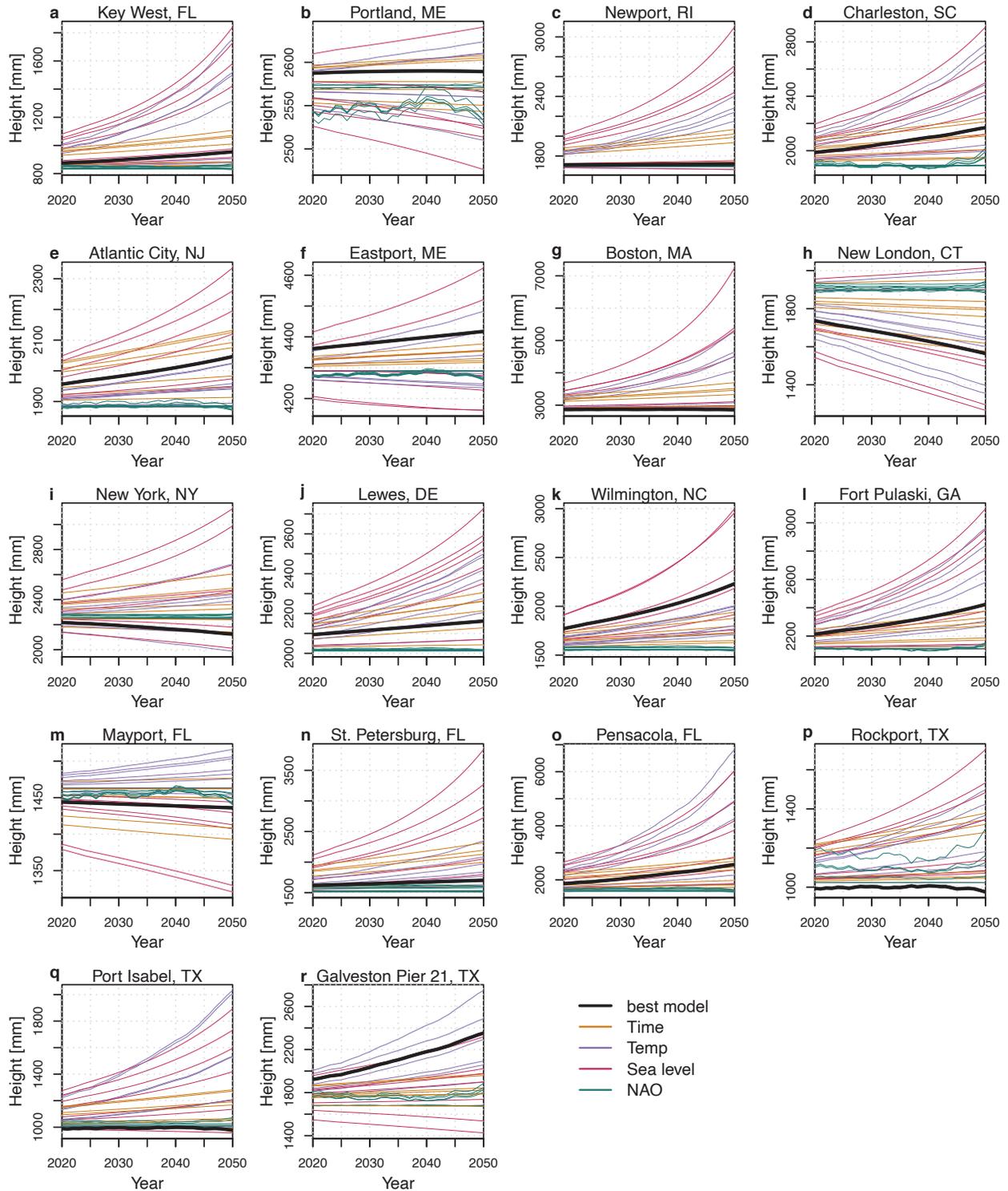

***Supplementary Figure 9.*** *Projected 50-year return levels for the long-data sites. Uses a GPD model and a centered 21-year moving window average. The best-fitting model is selected by minimizing the negative log-likelihood and is shown in black. The four candidate models using as the covariate time series are shown with the same colored lines: time (orange), temperature (purple), sea level (red) and NAO index (green). There is no visual distinction in each panel among the eight candidate model structures within the set using each covariate.*

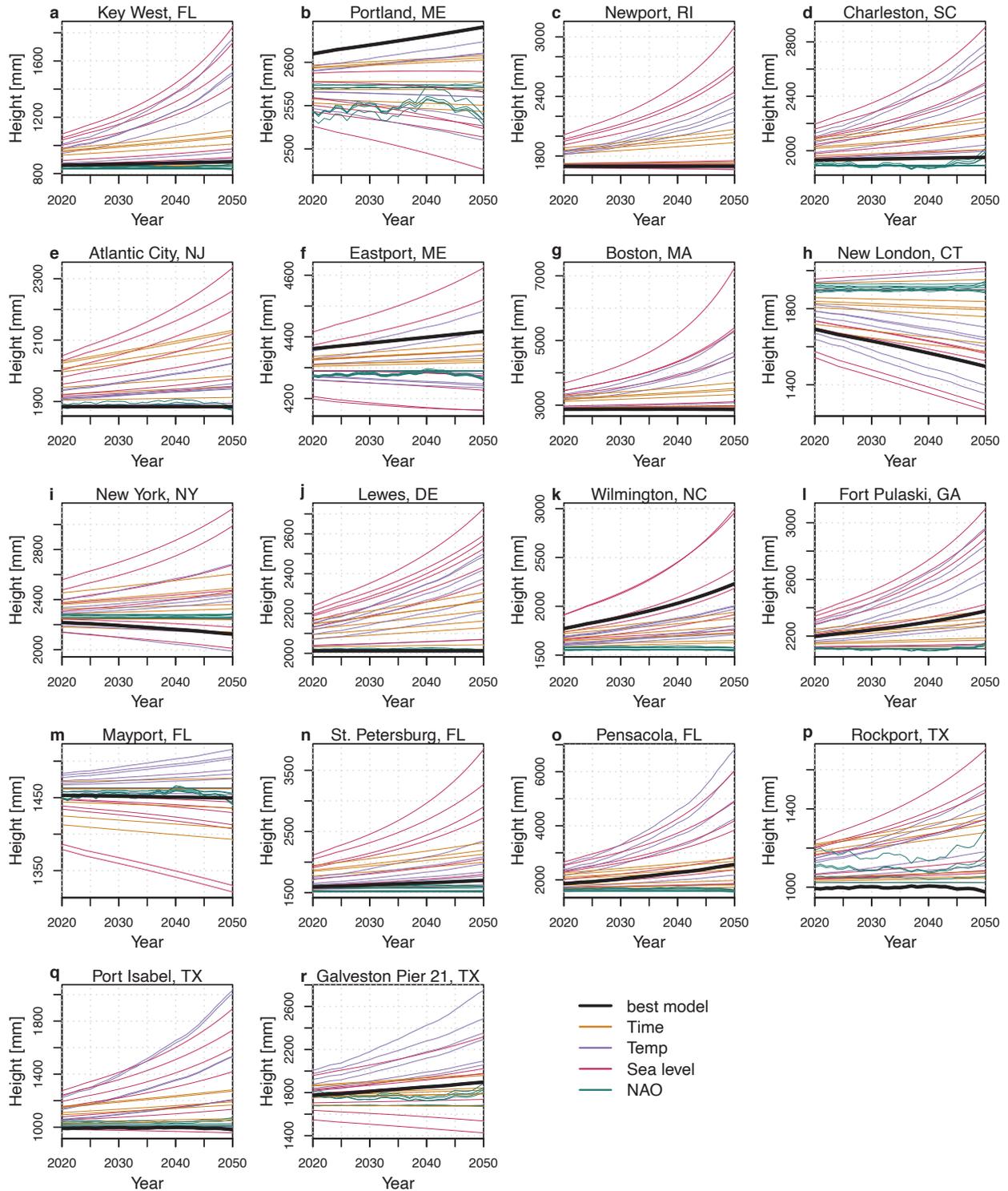

***Supplementary Figure 10.*** *Projected 50-year return levels for the long-data sites. Uses a GPD model and a centered 21-year moving window average. The best-fitting model is selected by minimizing the Akaike information criterion and is shown in black. The four candidate models using as the covariate time series are shown with the same colored lines: time (orange), temperature (purple), sea level (red) and NAO index (green). There is no visual distinction in each panel among the eight candidate model structures within the set using each covariate.*

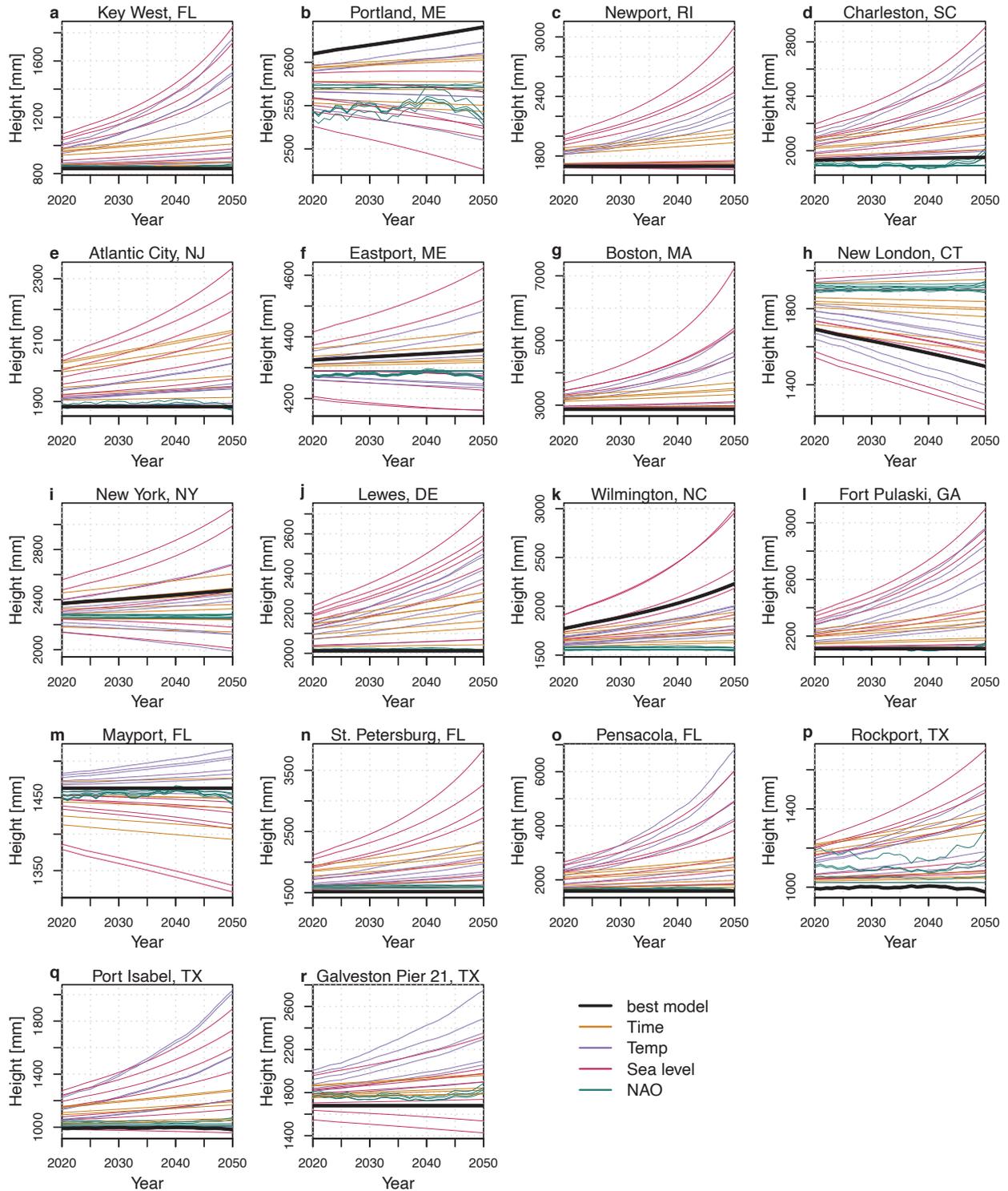

*Supplementary Figure 11. Projected 50-year return levels for the long-data sites. Uses a GPD model and a centered 21-year moving window average. The best-fitting model is selected by minimizing the Bayesian information criterion and is shown in black. The four candidate models using as the covariate time series are shown with the same colored lines: time (orange), temperature (purple), sea level (red) and NAO index (green). There is no visual distinction in each panel among the eight candidate model structures within the set using each covariate.*

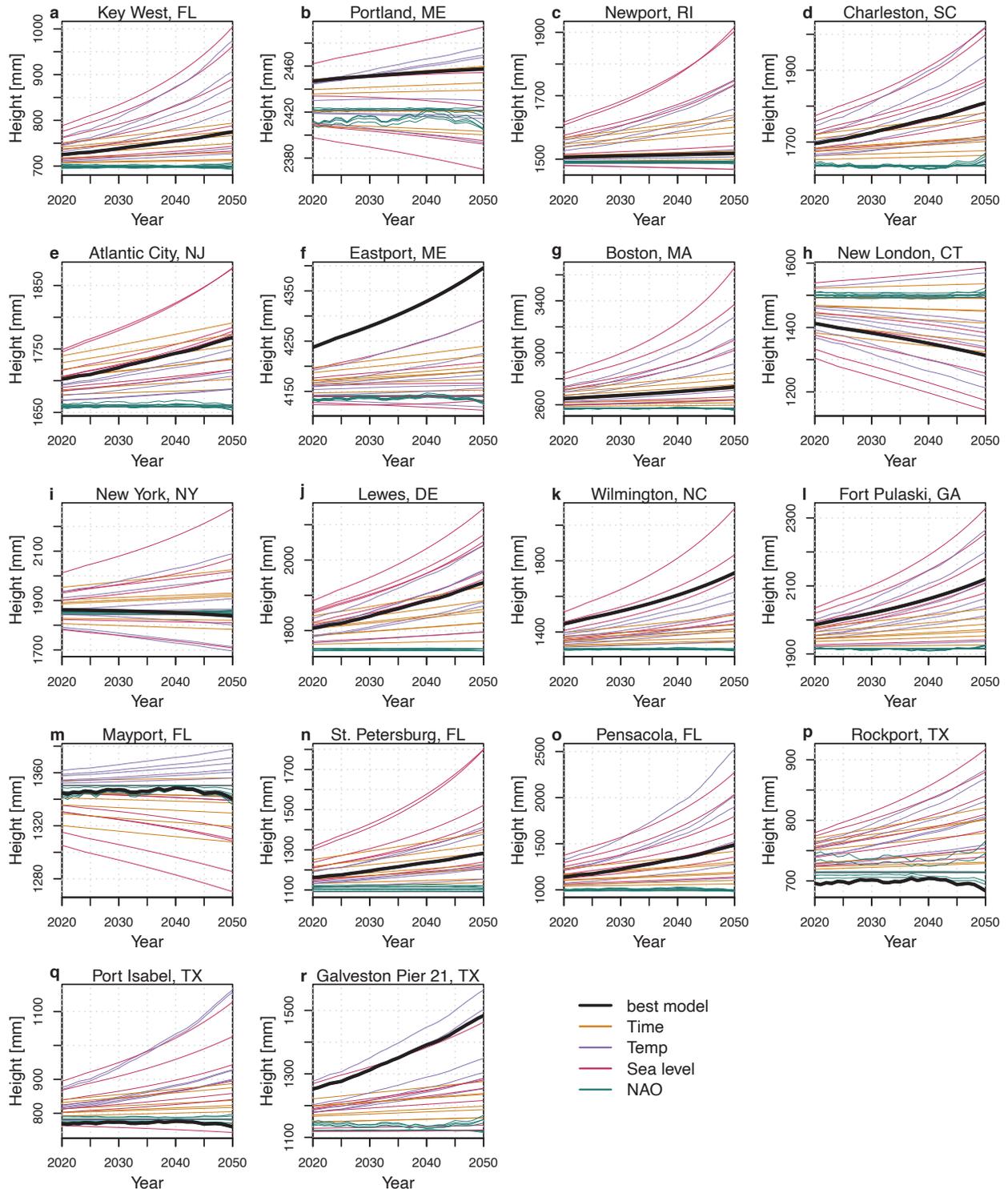

*Supplementary Figure 12. Projected 10-year return levels for the long-data sites. Uses a GPD model and a centered 21-year moving window average. The best-fitting model is selected by minimizing the negative log-posterior score and is shown in black. The four candidate models using as the covariate time series are shown with the same colored lines: time (orange), temperature (purple), sea level (red) and NAO index (green). There is no visual distinction in each panel among the eight candidate model structures within the set using each covariate.*

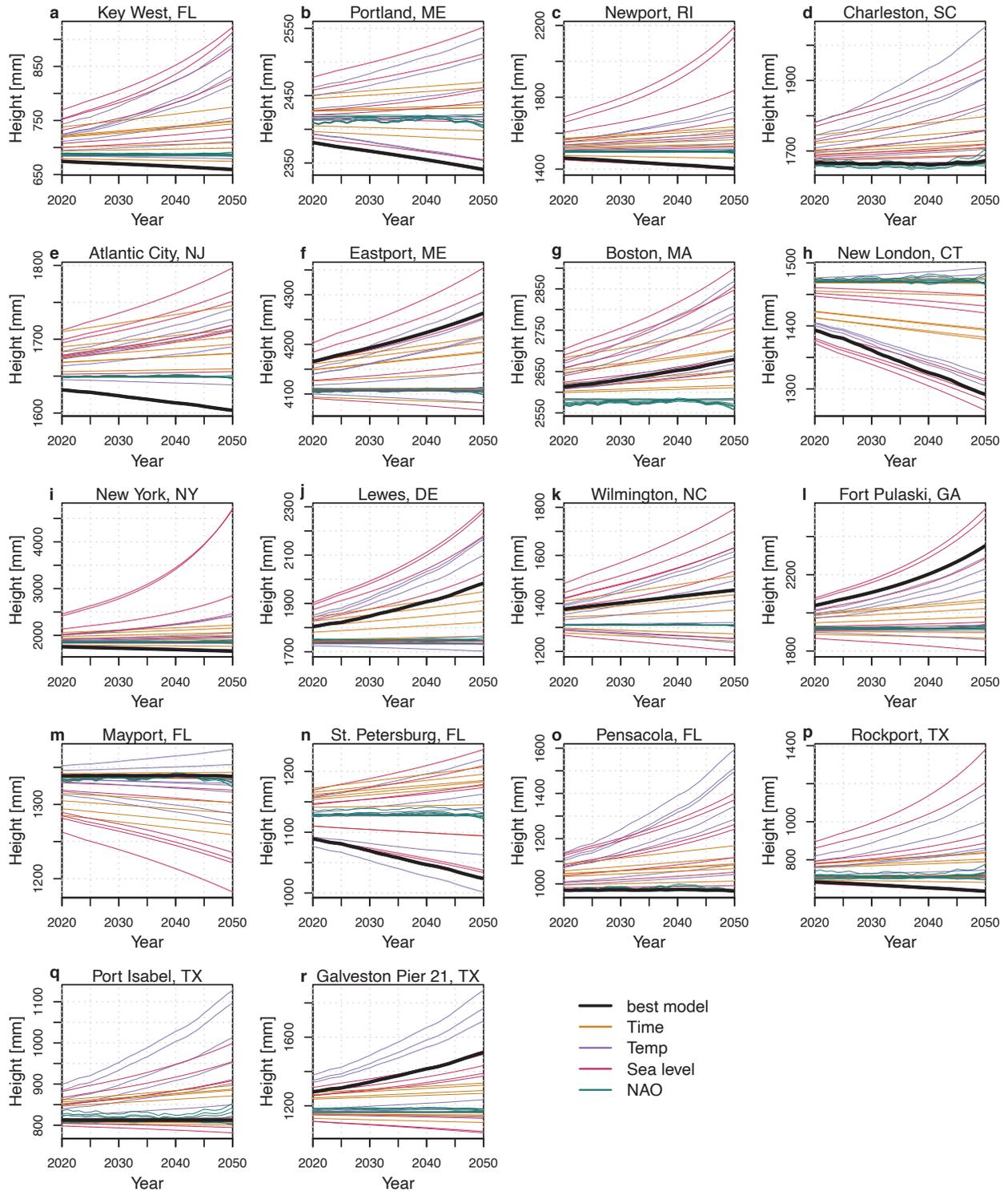

***Supplementary Figure 13.*** *Projected 10-year return levels for the long-data sites. Uses a GEV model and a centered 21-year moving window average. The best-fitting model is selected by minimizing the negative log-posterior score and is shown in black. The four candidate models using as the covariate time series are shown with the same colored lines: time (orange), temperature (purple), sea level (red) and NAO index (green). There is no visual distinction in each panel among the eight candidate model structures within the set using each covariate.*

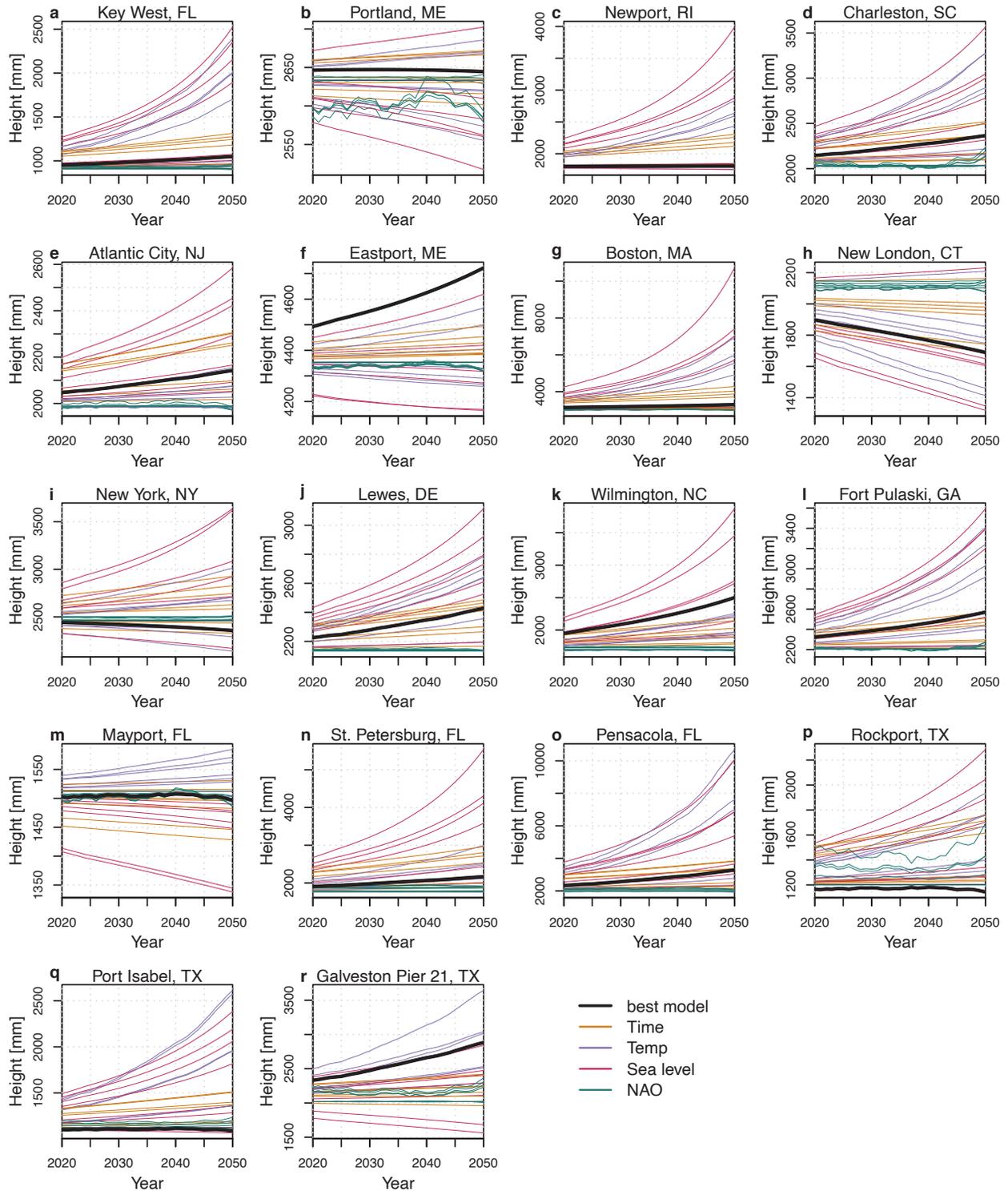

***Supplementary Figure 14.*** *Projected 100-year return levels for the long-data sites. Uses a GPD model and a centered 21-year moving window average. The best-fitting model is selected by minimizing the negative log-posterior score and is shown in black. The four candidate models using as the covariate time series are shown with the same colored lines: time (orange), temperature (purple), sea level (red) and NAO index (green). There is no visual distinction in each panel among the eight candidate model structures within the set using each covariate.*

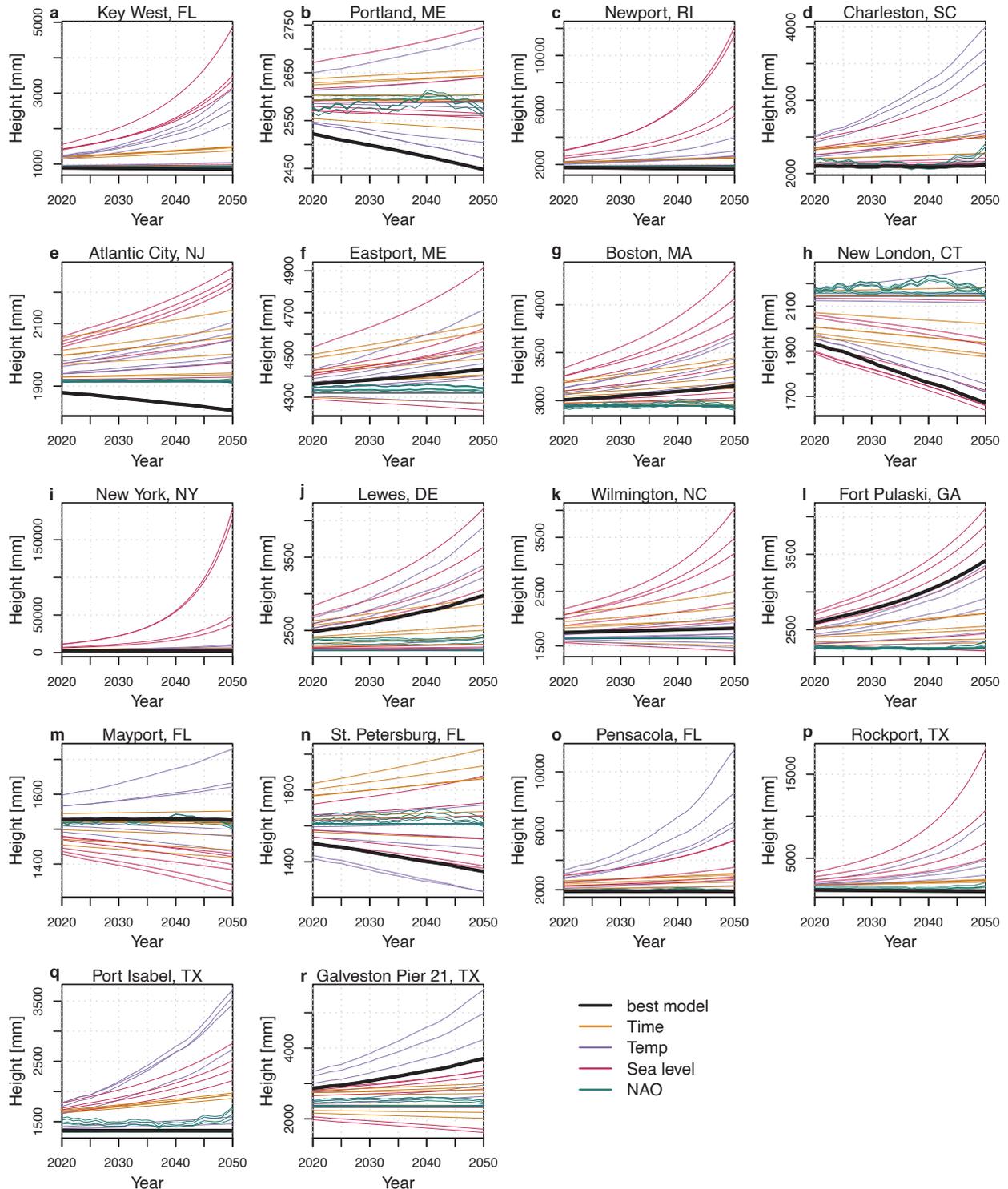

***Supplementary Figure 15.*** *Projected 100-year return levels for the long-data sites. Uses a GEV model and a centered 21-year moving window average. The best-fitting model is selected by minimizing the negative log-posterior score and is shown in black. The four candidate models using as the covariate time series are shown with the same colored lines: time (orange), temperature (purple), sea level (red) and NAO index (green). There is no visual distinction in each panel among the eight candidate model structures within the set using each covariate.*

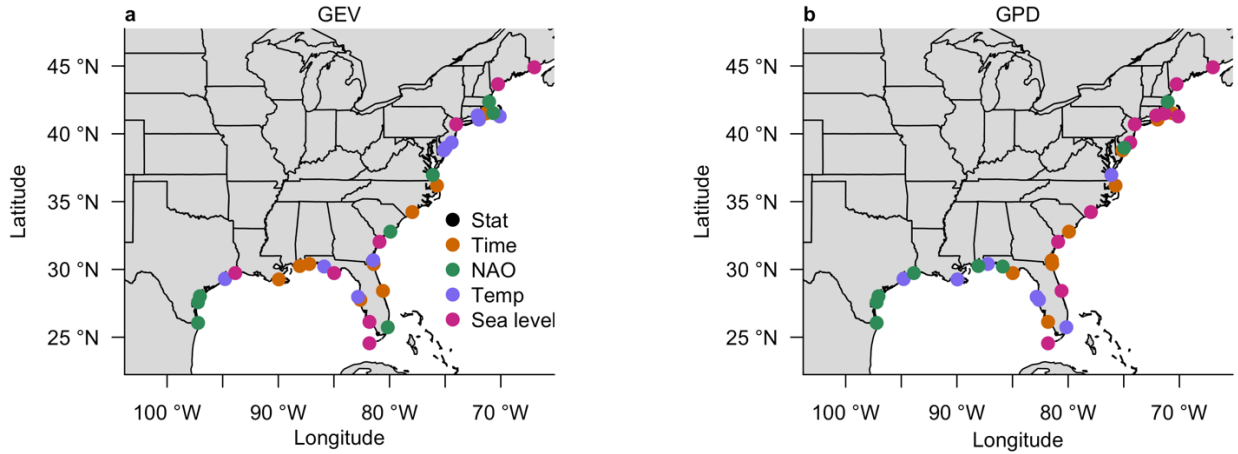

***Supplementary Figure 16.*** *Geographical distribution of the preferred covariate time series. Best-fit models are selected using negative log-likelihood to evaluate goodness-of-fit, assuming a GEV model (a) or a GPD one (b). Only extreme value statistical models with 0 (stationary) or 1 nonstationary parameter are considered; the shape parameter (ξ) is excluded.*

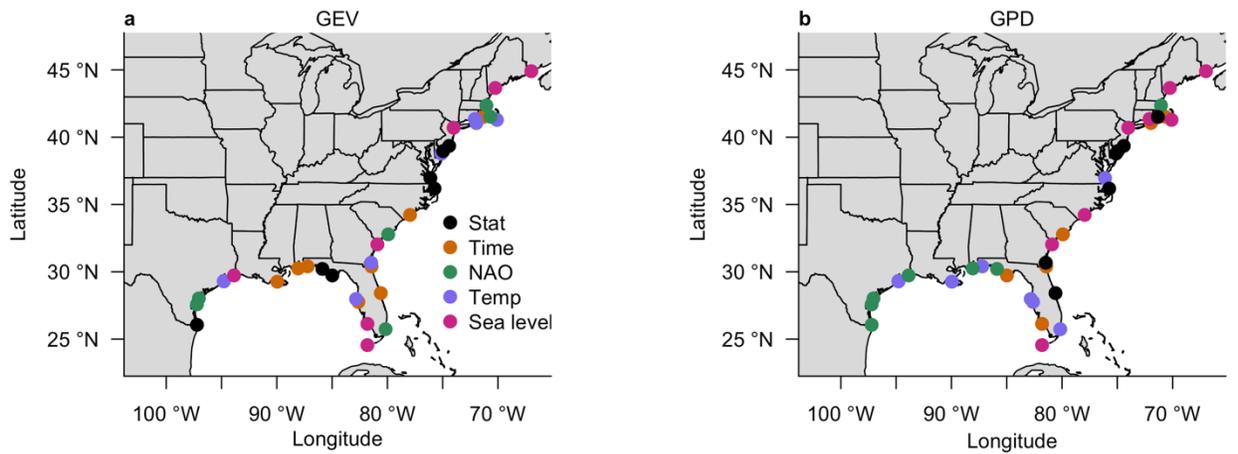

***Supplementary Figure 17.*** *Geographical distribution of the preferred covariate time series. Best-fit models are selected using Akaike information criterion to evaluate goodness-of-fit, assuming a GEV model (a) or a GPD one (b). Only extreme value statistical models with 0 (stationary) or 1 nonstationary parameter are considered; the shape parameter (ξ) is excluded.*

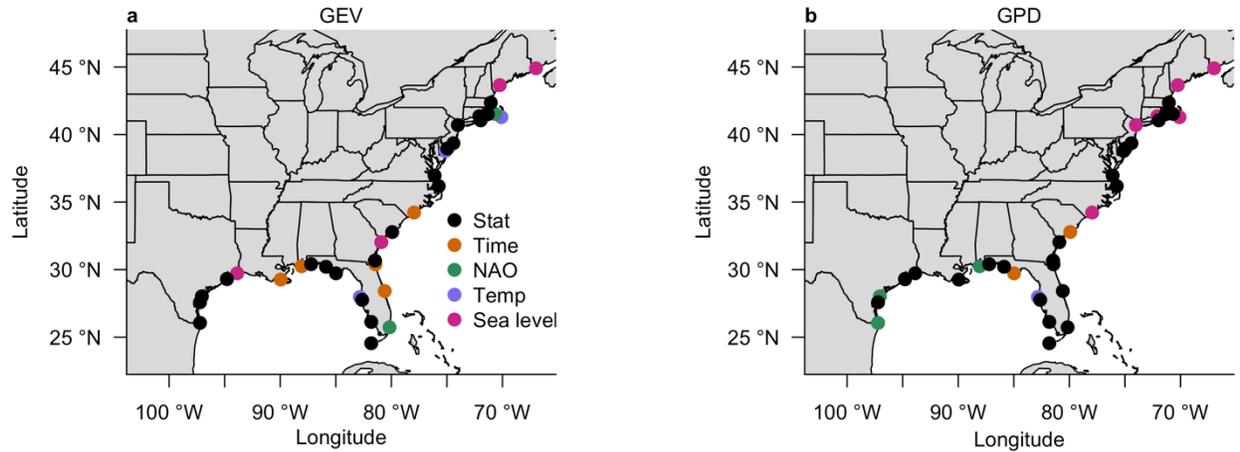

***Supplementary Figure 18.*** *Geographical distribution of the preferred covariate time series. Best-fit models are selected using Bayesian information criterion to evaluate goodness-of-fit, assuming a GEV model (a) or a GPD one (b). Only extreme value statistical models with 0 (stationary) or 1 nonstationary parameter are considered; the shape parameter ($\xi$) is excluded.*

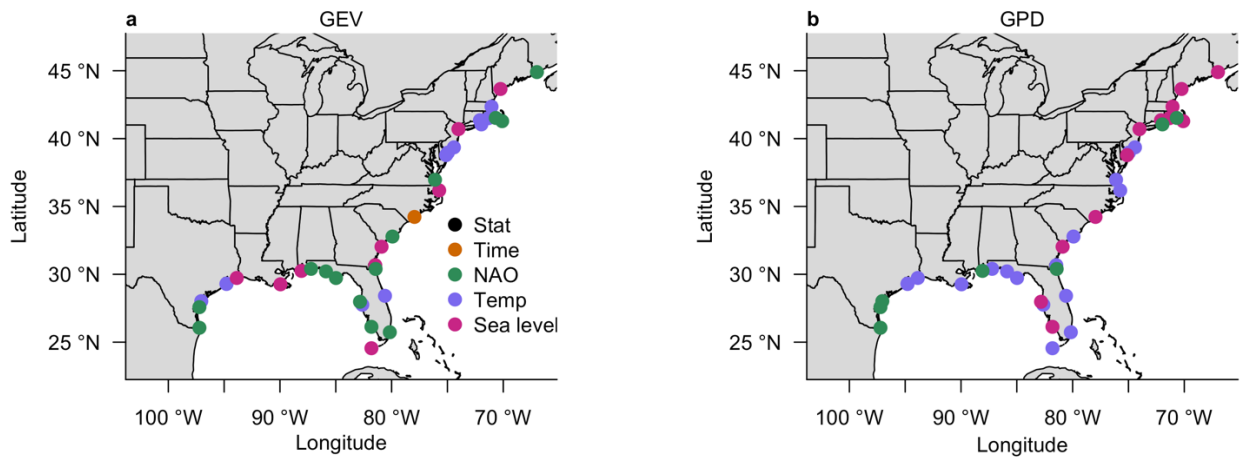

***Supplementary Figure 19.*** *Geographical distribution of the preferred covariate time series. Best-fit models are selected using negative log-posterior score to evaluate goodness-of-fit, assuming a GEV model (a) or a GPD one (b). Only extreme value statistical models with 0 (stationary) or 1 nonstationary parameter are considered; the shape parameter ($\xi$) is excluded.*

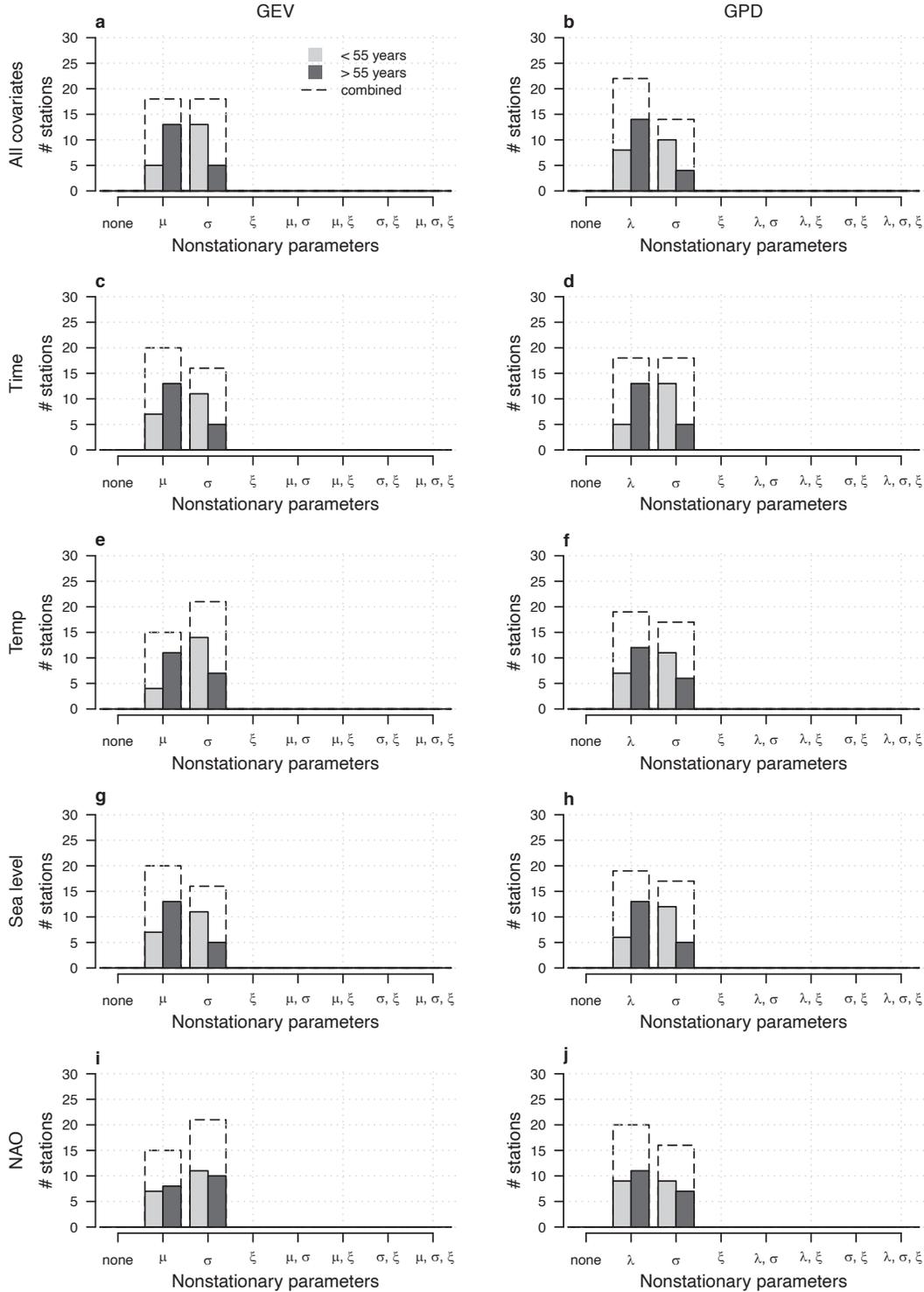

*Supplementary Figure 20. Frequency graphs of model choice, only considering extreme value statistical models with 0 (stationary) or 1 nonstationary parameter, excluding the shape parameter ($\xi$). Aggregates over all 36 tide gauge stations, and over all covariates (top row), or considering each covariate individually (time, second row; temperature, third row; sea level fourth row; and NAO index, fifth row). Left column corresponds to consideration of the eight GEV model structures and right column corresponds to the eight GPD models. Sites are separated into long (>55 years, dark gray) and short (<55 years, light gray) data record lengths. Negative log-likelihood is used to evaluate model goodness-of-fit.*

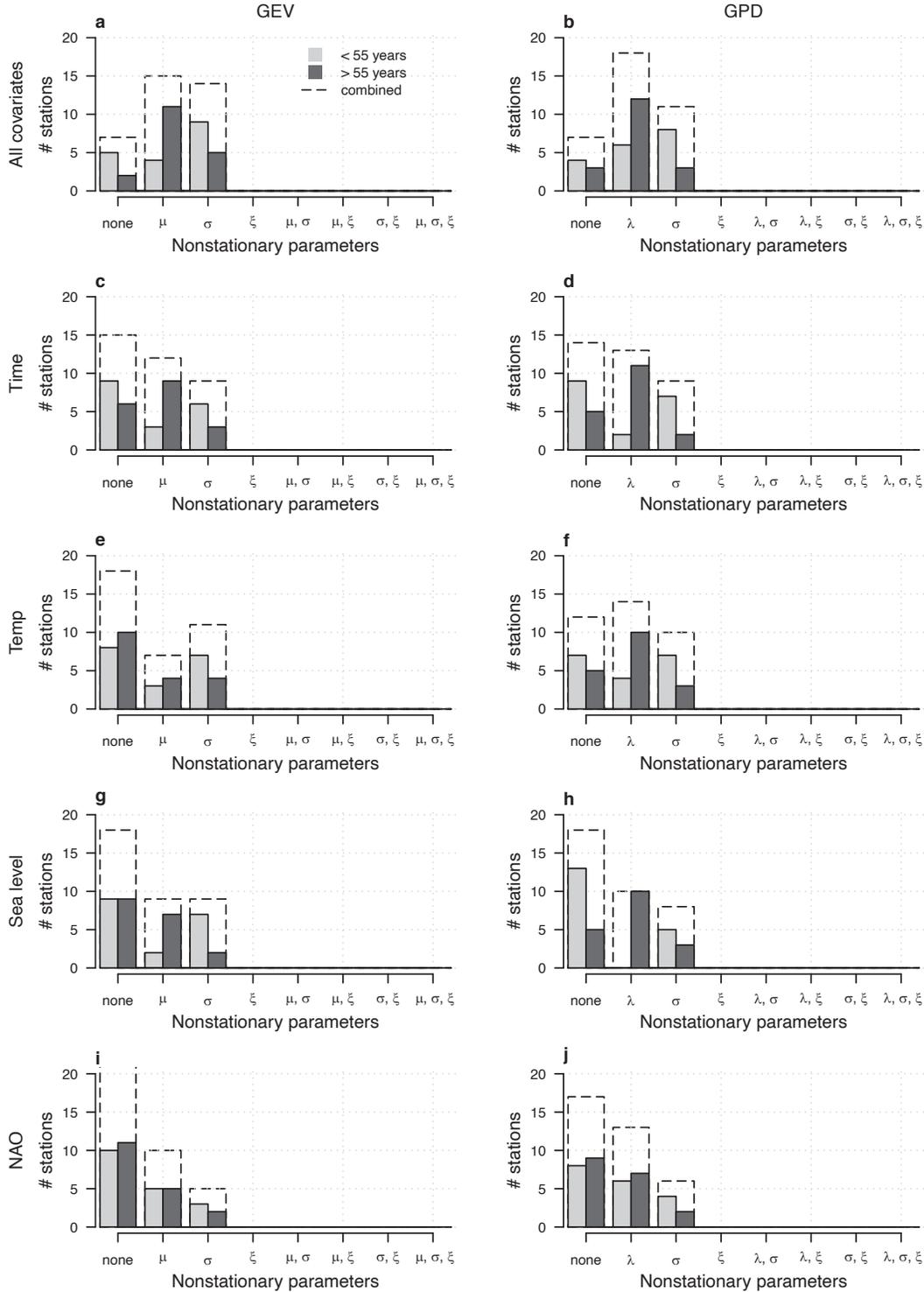

***Supplementary Figure 21.*** *Frequency graphs of model choice, only considering extreme value statistical models with 0 (stationary) or 1 nonstationary parameter, excluding the shape parameter ($\xi$). Aggregates over all 36 tide gauge stations, and over all covariates (top row), or considering each covariate individually (time, second row; temperature, third row; sea level fourth row; and NAO index, fifth row). Left column corresponds to consideration of the eight GEV model structures and right column corresponds to the eight GPD models. Sites are separated into long (>55 years, dark gray) and short (<55 years, light gray) data record lengths. Akaike information criterion is used to evaluate model goodness-of-fit.*

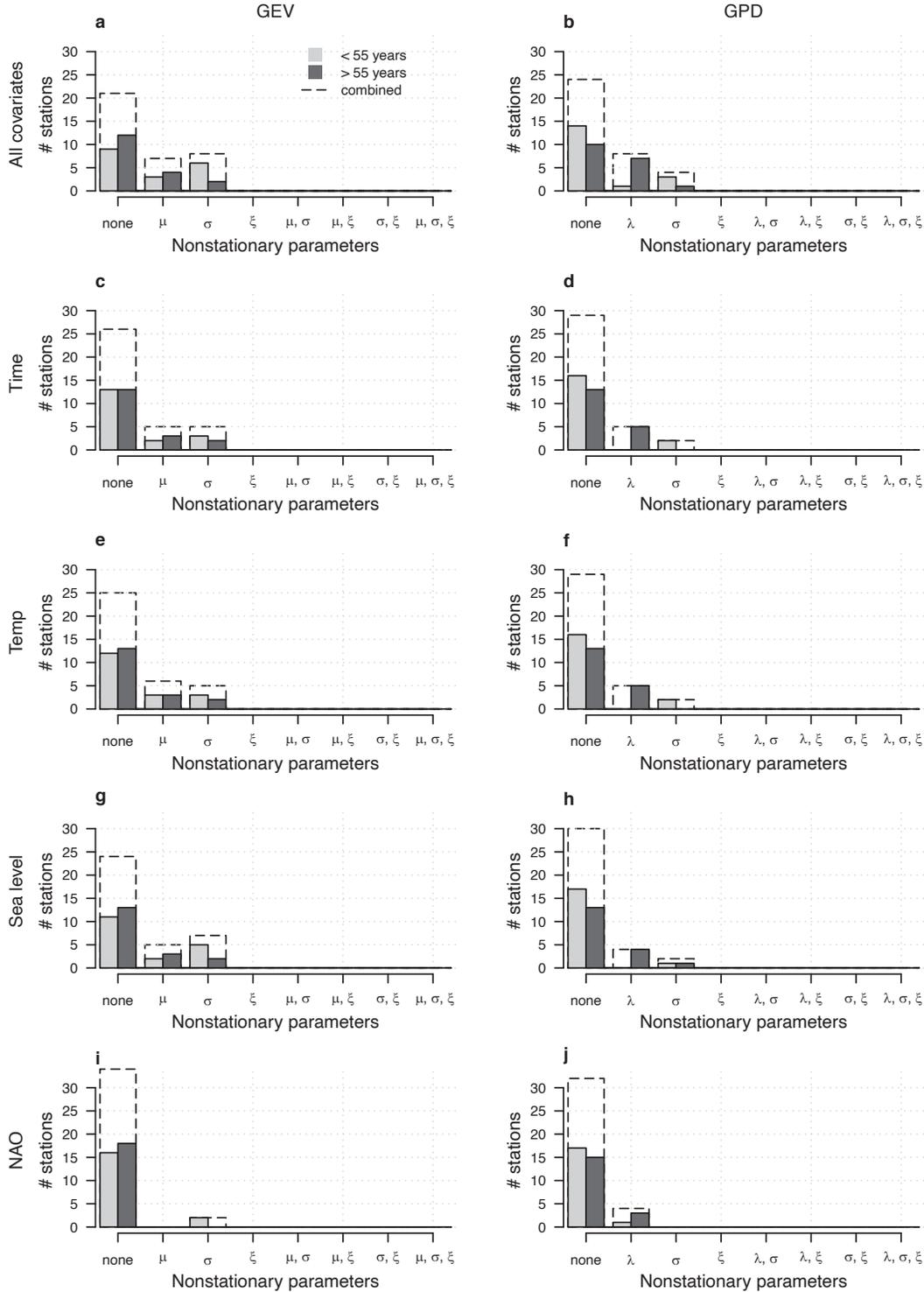

***Supplementary Figure 22.*** *Frequency graphs of model choice, only considering extreme value statistical models with 0 (stationary) or 1 nonstationary parameter, excluding the shape parameter ($\xi$). Aggregates over all 36 tide gauge stations, and over all covariates (top row), or considering each covariate individually (time, second row; temperature, third row; sea level fourth row; and NAO index, fifth row). Left column corresponds to consideration of the eight GEV model structures and right column corresponds to the eight GPD models. Sites are separated into long (>55 years, dark gray) and short (<55 years, light gray) data record lengths. Bayesian information criterion is used to evaluate model goodness-of-fit.*

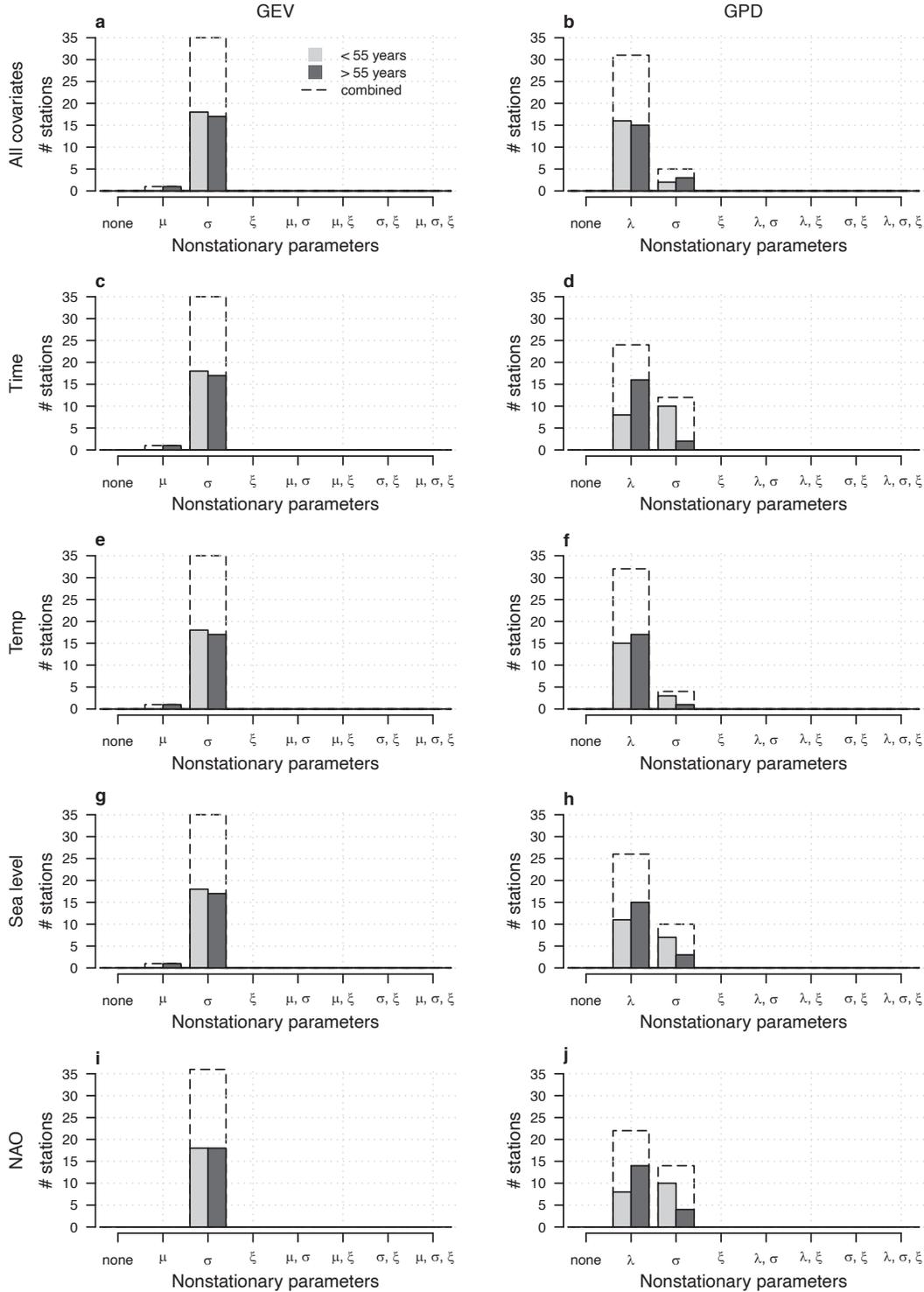

***Supplementary Figure 23.*** *Frequency graphs of model choice, only considering extreme value statistical models with 0 (stationary) or 1 nonstationary parameter, excluding the shape parameter ($\xi$). Aggregates over all 36 tide gauge stations, and over all covariates (top row), or considering each covariate individually (time, second row; temperature, third row; sea level fourth row; and NAO index, fifth row). Left column corresponds to consideration of the eight GEV model structures and right column corresponds to the eight GPD models. Sites are separated into long (>55 years, dark gray) and short (<55 years, light gray) data record lengths. Negative log-posterior score is used to evaluate model goodness-of-fit.*